\documentclass[final,3p]{elsarticle}
\usepackage{parskip}
\usepackage{amssymb}
\usepackage{amsmath}
\usepackage{mathrsfs}
\usepackage{mathtools}
\usepackage[pdfpagelabels]{hyperref}
\usepackage{pifont}
\usepackage{fontawesome}
\usepackage{stmaryrd}
\usepackage{enumitem}
\usepackage{multirow}
\usepackage{etoolbox}
\usepackage{graphicx}
\usepackage{gensymb}
\usepackage{amsthm}
\usepackage{centernot}
\usepackage[dvipsnames]{xcolor}
\usepackage{tikz}
\usepackage{tkz-euclide}
\usetikzlibrary{decorations.shapes}
\usepackage{algorithm}
\usepackage{algpseudocode}
\usepackage{appendix}
\usetikzlibrary{arrows,shapes.geometric}
\usetikzlibrary{calc}

\hypersetup{
	colorlinks=true,
	linkcolor=blue,     
	urlcolor=cyan,
}

\journal{Theoretical Computer Science}
\biboptions{sort&compress}

\newcommand\R{\mathbb{R}}

\newcommand\Q{\mathbb{Q}}
\newcommand\N{\mathbb{N}}

\newcommand\K{\mathcal{K}}

\newcommand\g{\mathtt{g}}
\newcommand\G{\mathtt{G}}

\newcommand\IF{\text{if }}

\newcommand\until{\mathcal{U}}
\newcommand\always{\square}
\newcommand\eventually{\lozenge}

\newcommand\limplies{\rightarrow}

\DeclareMathOperator*{\Int}{int}
\DeclareMathOperator*{\Ext}{ext}
\DeclareMathOperator*{\Iso}{iso}
\DeclareMathOperator*{\var}{var}
\DeclareMathOperator*{\Bd}{bd}
\DeclareMathOperator*{\Cl}{cl}

\newcommand\LTL{ \mathrm{LTL}}

\newtheorem{proposition}{Proposition}
\newtheorem{theorem}{Theorem}
\newtheorem{lemma}{Lemma}
\newtheorem{corollary}{Corollary}
\newtheorem{assumption}{Assumption}
\theoremstyle{definition}
\newtheorem{definition}{Definition}

\newtheorem{problem}{Problem}

\theoremstyle{remark}
\newtheorem{remark}{Remark}

\renewcommand{\theenumi}{\thedefinition.\alph{enumi}}

\begin{document}
	
	\begin{frontmatter}
		
		\title{Sampling Polynomial Trajectories for LTL Verification}
		
		\author[a]{Daniel Selvaratnam\corref{cor1}}
		\cortext[cor1]{Corresponding author}
		\ead{selvaratnamd@unimelb.edu.au}
		 
		\author[a]{Michael Cantoni}
		\author[a]{J. M.  Davoren}
		\author[b]{Iman Shames}

		\address[a]{Department of Electrical and Electronic Engineering, University of Melbourne, Parkville, VIC 3010, Australia}
	             
		\address[b]{College of Engineering and Computer Science, Australian National University, Canberra, ACT 2600, Australia}

		\begin{abstract}
			This paper concerns the verification of continuous-time polynomial spline trajectories against linear temporal logic specifications (LTL without `next'). Each atomic proposition is assumed to represent a state space region described by a multivariate polynomial inequality. The proposed approach samples a trajectory strategically, to capture every one of its region transitions. This yields a discrete word called a trace, which is amenable to established formal methods for path checking. The original continuous-time trajectory is shown to satisfy the specification if and only if its trace does. General topological conditions on the sample points are derived that ensure a trace is recorded for arbitrary continuous paths, given arbitrary region descriptions. Using techniques from computer algebra, a trace generation algorithm is developed to satisfy these conditions when the path and region boundaries are defined by polynomials. The proposed \textsc{PolyTrace} algorithm has polynomial complexity in the number of atomic propositions, and is guaranteed to produce a trace of any polynomial path. Its performance is demonstrated via numerical examples and a case study from robotics. 	
		\end{abstract}
	
	
\end{frontmatter}

\section{Introduction}
Path checking~\cite{basinOptimalProofsLinear2018} is about testing whether a particular signal (or word) satisfies a given formal specification. By contrast, the aim in model checking~\cite{baierPrinciplesModelChecking2008} is to establish whether all outputs from a model satisfy the specification. Path checking is generally much easier than model checking~\cite{markeyModelCheckingPath2003}, particularly for infinite state models, for which the model checking problem is usually undecidable. Hence runtime verification, a popular approach to formal system verification~\cite{havelundRuntimeVerification172018}, abandons model checking in favour of path checking, at the expense of exhaustive coverage. Path checking algorithms for discrete words are well studied~\cite{markeyModelCheckingPath2003,basinOptimalProofsLinear2018,bauerRuntimeVerificationLTL2011}. We focus here on path checking continuous-time signals. Signal Temporal Logic (STL) is introduced in \cite{malerMonitoringTemporalProperties2004} as part of a framework for path checking the outputs of continuous-time and hybrid systems. STL is defined there as a fragment of Metric Temporal Logic (MTL)~\cite{koymansSpecifyingRealtimeProperties1990} in which each atomic proposition is associated with a region of the state-space. A procedure is then developed in \cite{malerMonitoringTemporalProperties2004} for path checking continuous-time signals against STL specifications with bounded temporal operators. This procedure relies on sampling the continuous-time signal in a manner assumed dense enough to capture all signal transitions between the relevant state space regions~\cite{malerMonitoringTemporalProperties2004}. Satisfying this assumption is a decidedly non-trivial problem. Here, we present a sampling strategy that guarantees it, assuming that the signal to be verified is a polynomial spline, and that the state space regions of interest are semi-algebraic sets. 

A useful comparison between path checking and model checking is provided in \cite[Section 1.1]{bauerRuntimeVerificationLTL2011}. A recurring issue relates to word length: most temporal logics are interpreted over infinite words, but we are practically constrained to store and process only finite words. In discrete-time, one way to resolve this is by restricting attention to \emph{lasso words}, which are infinite words of the form $\alpha\cdot\beta^\omega$, for finite $\alpha,\beta$. Path checking algorithms for such words are devloped in \cite{markeyModelCheckingPath2003, basinOptimalProofsLinear2018}. Other ways to tackle the finite length issue for discrete words are discussed in \cite{eisnerReasoningTemporalLogic2003}. 
In continuous time, a different approach has been to only path check fragments of MTL with bounded temporal operators~\cite{malerMonitoringTemporalProperties2004, fainekosRobustnessTemporalLogic2009, donzeRobustSatisfactionTemporal2010}, the satisfaction of which are fully determined by signal prefixes over finite time horizons. In contrast to this, we focus on Linear Temporal Logic (LTL) specifications, which can be considered fragments of MTL restricted to unbounded temporal operators. This choice is motivated by the control systems literature~\cite{kloetzerFullyAutomatedFramework2008, liuSynthesisReactiveSwitching2013, wongpiromsarnAutomataTheoryMeets2016}, in which controllers for continuous-time systems are synthesized to meet closed-loop LTL specifications. We address the verification problem, not the synthesis problem. The proposed sampling strategy generates a \emph{trace} of the continuous-time signal to be verified, in line with the definition of such from \cite{wongpiromsarnAutomataTheoryMeets2016}. Intuitively, a trace is a discrete word that captures every region transition taken by the continuous-time signal. In Section \ref{sec:samplingForTraces} we formalise the relationship between traces and the \emph{timed state sequences} (TSSs) of \cite{alurBenefitsRelaxingPunctuality1996}. We also identify a class of continuous-time signals having lasso word traces, and argue these are the only continuous-time signals with unbounded domains that are practically amenable to LTL path checking. 

The notion of a robustness margin has proved valuable for path checking continuous-time signals by sampling. Intuitively, a robustness margin quantifies how robustly a formal specification is satisfied or how severely it is violated. In order to conclude that a continuous-time signal satisfies the specification, its sampled version is typically required to satisfy a corresponding discrete-time specification, with a sufficiently large positive robustness margin. Such an approach is proposed in \cite{fainekosRobustnessTemporalLogic2009}, in which MTL semantics that can be interpreted over both discrete and continuous-time signals are developed. The signals are assumed to take values in a metric space, and a robustness margin is defined based on the metric. Moreover, the continuous-time signals are assumed to satisfy a generalized global Lipschitz condition~\cite[Assumption 35]{fainekosRobustnessTemporalLogic2009}, and an upper bound on the sampling time is also required. Then, continuous-time satisfaction can be inferred from sufficiently robust discrete-time satisfaction. The same approach is adopted in \cite{babinRefinementProofBased2015}. The drawback is that the required sampling frequency may not be known \emph{a priori}. Thus, if the robustness margin is not large enough for a particular sampled signal, then violation of the specification for the continuous-time signal cannot be inferred. The test can only be repeated with a different sampling strategy, which may again yield an inconclusive result. It is indeed possible for a continuous-time signal to satisfy or violate a specification with zero robustness, in which case the test never yields a conclusive result. A distinctive feature of our approach is that it does not rely on robustness margins at all. Since our algorithm is guaranteed to produce a trace, the trace can be checked against any LTL specification to always yield a conclusive result, even when the robustness margin is zero. To generate the trace, our sampling strategy does not depend on the specification imposed, neither does it rely on an upper bound between sampling times, thereby avoiding unnecessary sampling over intervals in which no transitions occur. This comes at the cost of considering a narrower class of continuous signals and regions of interest, specifically, those that are fully characterised by polynomials. However, such polynomial signals need not satisfy \cite[Assumption 35]{fainekosRobustnessTemporalLogic2009}. 

Building on \cite{fainekosRobustnessTemporalLogic2009}, subsequent works have estimated the robustness margin of a continuous-time signal from its samples by computing the robustness margin of its reconstructed version. Linear interpolation between sample points is assumed in \cite{donzeRobustSatisfactionTemporal2010, donzeEfficientRobustMonitoring2013}. This approach is generalised in \cite{abbasTemporalLogicRobustness2019} to allow for reconstruction via polynomial spline interpolation. Error bounds between the robustness margin of the original signal and the reconstructed spline are also derived. Our approach relates to the computation of the spline robustness margins in \cite{abbasTemporalLogicRobustness2019}, but with some key differences. Firstly, we focus only on evaluating the boolean satisfaction of a specification by a signal, without attempting to compute its robustness margin. This allows us to verify edge cases that have zero (or almost zero) robustness, which would otherwise yield an inconclusive result. Secondly, \cite{abbasTemporalLogicRobustness2019} relies on numerical root finding to find inflection and intersection points of polynomials, without specifying a particular algorithm. Numerical root finding can be expensive. It also introduces errors, because, in general, polynomial roots are irrational, and cannot be found exactly for polynomials of degree five and higher. The choice of root finding algorithm is important, because this determines the overall complexity of the algorithm, and some root finders are not guaranteed to find every root (e.g. Newton's method~\cite[Chapter 9]{vonzurgathenModernComputerAlgebra2013}). Polynomial roots are also central to our approach, however we adopt root isolation techniques from computer algebra, which rely on symbolic computation over the rationals, and come with theoretical guarantees. The roots do not actually need to be found, only isolated. As a result, no numerical errors are introduced during our computations. Finally, the procedure in \cite{abbasTemporalLogicRobustness2019} appears to be sound, but it is stated without proof. A key contribution of our work is to provide a rigorous proof of correctness.

The restriction of scope to polynomial signals and semi-algebraic regions of interest may at first appear severe, however this class of verification problems is of practical significance. The use of polynomial splines for signal reconstruction in \cite{abbasTemporalLogicRobustness2019} has already been mentioned. Robotic motion planning is also frequently solved by constructing a polynomial spline through a sequence of waypoints~\cite{burkeFastSplineTrajectory2021, mercySplineBasedMotionPlanning2018, lauKinodynamicMotionPlanning2009, shillerDynamicMotionPlanning1991}. Our algorithm can be used to verify a proposed motion plan against mission requirements expressed in LTL. This is illustrated by a case study in Section \ref{sec:caseStudy}. A special case of this problem, focused only on detecting collisions, is treated in \cite{yoonSplinebasedRRTUsing2018}. Collision detection is also the focus of \cite{weinContinuousPathVerification2004}, which assumes rational rather than polynomial spline paths. Semi-algebraic regions include half-spaces, circles, ellipses, hyperbolas, and finite intersections and unions thereof. These suffice to approximate obstacles or target regions to any desired degree of fidelity. Polynomials are of theoretical interest as well, because the celebrated Weierstrass Approximation Theorem assures us that any continuous function can be approximated uniformly and arbitrarily closely by a polynomial. Polynomials, then, are worthy of special attention, and our work exploits their properties for verification, by using the appropriate tools from computer algebra. 

\emph{Overview:} At its core, our algorithm involves sampling a continuous-time trajectory on either side of (and sometimes at) every crossing of the boundary of a region of interest. Since boundaries are a topological property, Section \ref{sec:topology} provides some basic definitions and results from point-set topology, which are used along the way. We also rely on the preliminaries from algebra reviewed in Section \ref{sec:algebra} to exploit the algebraic properties of polynomials for their verification. The general problem we consider is formulated precisely in Section \ref{sec:problem}, which reviews the continuous-time semantics of MITL, and then presents LTL without next as a fragment of MITL. Section \ref{sec:samplingTheory} lays the theoretical foundations of the paper. A distinction is made between two types of continuous signals: paths and trajectories, the former having a finite trace, and the latter an infinite one. This enables us to verify the class of trajectories that can be decomposed into a finite number of paths, because such trajectories have lasso word traces. Sections \ref{sec:paths} and \ref{sec:traceOperations} address the decomposition of trajectories into paths, and the related construction of trajectory traces from path traces. The first main result of the paper is Theorem \ref{thm:topological} in Section \ref{sec:abstract}, which presents topological conditions to ensure that samples are taken on either side of every boundary crossing, and when necessary, at the crossings themselves. These conditions are stated in terms of sampling functions, which are introduced in Section \ref{sec:samplingForTraces} to describe arbitrary sampling strategies. The theory developed in Section \ref{sec:samplingTheory} is applied in Section \ref{sec:concrete} to polynomial paths travelling through semi-algebraic regions of interest. Section \ref{sec:isolatedPointCharacterisation}, in particular, identifies the algebraic conditions that require samples to be taken exactly on a boundary. This precedes Theorem \ref{thm:polynomials}, which expresses the topological conditions of Theorem \ref{thm:topological} in terms of polynomial roots. Section \ref{sec:numerical} then constructs the \textsc{PolyTrace} algorithm to satisfy the conditions of Theorem \ref{thm:polynomials}. An important result is Lemma \ref{lem:sampleIsolated} (the Isolation Lemma), which permits a trajectory to be sampled exactly on a boundary, with only approximate knowledge of the crossing point. \textsc{PolyTrace} relies on root isolation algorithms and a root existence test from computer algebra, which are discussed in Sections \ref{sec:rootExistence} and \ref{sec:rootIsolation}. The final main result is Theorem \ref{thm:alg}, which establishes the correctness of \textsc{PolyTrace}. Numerical examples demonstrate the capabilities of the algorithm in Section \ref{sec:sim}, and a case study applies it to the verification of a robot motion plan. In particular, the case study imposes an LTL specification that can only be satisfied with zero robustness. Such specifications simply cannot be verified using existing methods, but yield readily to \textsc{PolyTrace}. Concluding remarks are then made in Section \ref{sec:conclusion}.
\section{Preliminaries} \label{sec:prelim}

\subsection{Notation} \label{sec:notation}
Let $\R$ denote the real numbers, $\Q$ the rational numbers, $\mathbb{Z}$ the integers, and $\mathbb{N} := \{0,1,...\}$ the natural numbers. For any $M \subset \N$, let $M^\ominus:= \{ m  \in M \mid m+1 \in M\}$. The subset relation is denoted by $\subset$, and the strict subset relation by $\subsetneq$. A set is \emph{countable} iff it is either finite or countably infinite. If $f:A \to B$, then let $f[X] \subset B$ denote the image of $X \subset A$, and $f^{-1}Y \subset A$ the preimage of $Y \subset B$.
The composition of $g:B \to C$ with $f:A \to B$ is denoted by $g \circ f:A \to C$. The restriction of $f$ to $D \subset A$ is denoted by $f {\restriction_D}$. For two integers $i \leq j$, let $a_{i:j} := (a_i, a_{i+1}, \hdots, a_{j-1}, a_j)$. For products taken over the empty set, define $ \prod_{i \in \emptyset} (\cdot) := 1$. 

\subsection{Topology} \label{sec:topology}
Let $X$ be a topological space. A set $U \subset X$ is a neighbourhood of $x \in X$ iff $U$ is open and $x \in U$. A topological space is \emph{connected} iff it is not the union of two of its disjoint open subsets \cite[\S 23]{munkresTopology2000}. 
A point $x \in X$ is an \emph{isolated point} of $X$ iff $\{x\}$ is open ~\cite[Section 1.3]{engelkingGeneralTopology1989}. Let $\Iso(X)$ denote the set of isolated points of $X$. Moreover, let $\Int(A)$ and $\Cl(A)$ denote the interior and closure of $A \subset X$, respectively. Then $\Bd(A):= \Cl(A) \setminus \Int(A)$ denotes the boundary of $A$, and $\Ext(A) := \Int(X \setminus A)$ the exterior of $A$. A point $x \in \Bd(A)$ iff every neighbourhood of $x$ intersects both $A$ and $X \setminus A$ \cite[Proposition 1.3.1]{engelkingGeneralTopology1989}. The result below is a corollary of this. 
\begin{proposition} \label{prop:bdX}
	For any topological space $X$, $\Bd(X) = \emptyset$. 
\end{proposition}
Any $S \subset X$ becomes a topological space when endowed with the subset topology. That is, $V \subset S$ is declared open in $S$ iff $V = S \cap U$ for some $U \subset X$ open in $X$. Then $S$ is referred to as a subspace of $X$. It follows from the definition of the subset topology that $s \in \Iso(S)$ iff there exists $U \subset X$, open in $X$, such that $U \cap S = \{s\}$.
We refer to $V \subset S$ as a $S$-neighbourhood of $s \in S$ iff $V$ is open in $S$ and $s \in V$.  

\begin{proposition} \label{prop:isoInBd}
	Let $A \subset X$, where $X$ is a topological space. If $X$ contains no isolated points, then
	$ \Iso(A) \subset \Bd(A). $ 
\end{proposition}
\begin{proof}
	Let $a \in \Iso(A) \subset A$. There then exists $U \subset X$ open in $X$ such that \begin{equation} U \cap A = \{a\}. \label{eq:singleton}\end{equation} 
	Choose any neighbourhood $V$ of $a$. Now $a \in U \cap V \neq \emptyset$. Furthermore, $U \cap V$ is open in $X$. If $U \cap V = \{a\}$, then $a \in \Iso(X)$. Assume $X$ contains no isolated points. Then $\{a\} \subsetneq U \cap V$, and there exists $x \in U \cap V$ such that $x \neq a$. If $x \in A$, then $x \in U \cap A$, which contradicts \eqref{eq:singleton}.  Thus $x \in X \setminus A$, and $x \in V$, so $V \cap (X \setminus A) \neq \emptyset$. Clearly also  $a \in V \cap A \neq \emptyset$. Thus $a \in \Bd(A)$.
\end{proof}

\begin{proposition} \label{prop:BdIntersection}
	Let $A, B \subset X$, where $X$ is a topological space. Then 
	$\Bd(A \cap B) \subset \Bd(A) \cup \Bd(B).$
\end{proposition}
\begin{proof}
	By \cite[Theorem 1.3.2, (iv)]{engelkingGeneralTopology1989}, $\Bd(A \cap B) \subset \big(\Bd(A) \cap \Cl(B)\big) \cup \big(\Bd(B) \cap \Cl(A)\big).$
\end{proof}
\begin{lemma} \label{lem:isoMinus}
	Let $A,B \subset X$, where $X$ is a topological space. Then
	$ \Iso(A \cap B) \setminus \Bd(B) \subset \Iso(A).$
\end{lemma}
\begin{proof}
	Suppose \begin{equation} s \in \Iso(A \cap B) \subset (A \cap B). \label{eq:isoInt} \end{equation} Then $V \cap A \cap B = \{s\}$, for some neighbourhood $V$ of $s$. Suppose in addition that $s \notin \Bd(B)$. Then \eqref{eq:isoInt} implies $s \in B \setminus \Bd(B) = \Int(B)$ \cite[Theorem 1.3.2, (i)]{engelkingGeneralTopology1989}. Thus, \begin{equation} U \subset B \label{eq:UsubB} \end{equation} for some neighbourhood $U$ of $s$. Observe that $V \cap U$ is also a neighbourhood of $s$. In particular, $s \in V \cap U$, and \eqref{eq:isoInt} then implies $s \in V \cap U \cap A$. Furthermore, $(V \cap U \cap A) \subset (V \cap B \cap A) = \{s\}$ by \eqref{eq:UsubB}, and therefore $V \cap U \cap A = \{s\}$, which implies $s \in \Iso(A)$. 
\end{proof}

Suppose $A \subset S \subset X$, and let $\Int(A|S),\ \Cl(A|S),\ \Bd(A|S),\ \Ext(A|S)$ respectively denote the interior, closure, boundary and exterior of $A$ relative to $S$ (i.e., with respect to the subset topology on $S$). 
\begin{remark} \label{rem:sub2main}
	Since $S$ is a topological space when endowed with the subset topology, Propositions \ref{prop:isoInBd}, \ref{prop:BdIntersection} and Lemma \ref{lem:isoMinus} all hold when $\Bd(\cdot)$ is replaced with $\Bd( \cdot|S)$, provided that $A,B \subset S\subset X$.
\end{remark}
We will assume the standard topology on $\R^n$, which in particular implies that $\R^n$ (and hence $\R$) have no isolated points.  
\begin{lemma} \label{lem:noCrossing}
	Let $U  \subset S \subset \R$, and suppose $I \cap U  \neq \emptyset$ for some interval $I \subset S$. Then $$ I \cap \Bd \left( U  \mid S \right) =\emptyset \implies I \subset U .$$
\end{lemma}
\begin{proof}
	Firstly, $S = \Int(U |S) \cup \Bd(U |S) \cup \Ext(U |S)$, and therefore
	\begin{equation} I =( I \cap \Int(U |S)) \cup (I \cap \Bd(U |S)) \cup (I \cap \Ext(U |S)). \label{eq:decompI} \end{equation}
	By definition of the subset topology, $I \cap \Int(U |S)$ and $I \cap \Ext(U |S)$ are open in $I$. Suppose $I \cap \Bd(U |S) = \emptyset$.  Since $U  \cap I \neq \emptyset$, this implies $I \cap \Int(U |S) \neq \emptyset$. If $I \cap \Ext(U |S) \neq \emptyset$, then \eqref{eq:decompI} implies $I$ is the union of two non-empty disjoint open subsets. However, $I$ is an interval and therefore connected, which is a contradiction. Thus $I \cap \Ext(U |S) = \emptyset$, and \eqref{eq:decompI} then implies $I \subset \Int(U |S) \subset U$. 
\end{proof}
\subsection{Algebra} \label{sec:algebra}
Let $R[x_1,...,x_n]$ denote the ring of polynomials in the variables $x_1,...,x_n$ with coefficients in $R$, where $R$ is a commutative ring with multiplicative identity \cite[Chapter 9]{dummitAbstractAlgebra2004}. Since $\R$ is an infinite integral domain, we do not distinguish between a polynomial $p \in \R[x_1,...,x_n]$ and the polynomial function $p:\R^n \to \R$ associated with it~\cite[Chapter 3, Section 5.1]{mignotteMathematicsComputerAlgebra1992}. Let $\deg(p)$ denote the degree of $p$. In what follows, we are particularly concerned with the Euclidean domain $\R[s]$ of univariate polynomials with real coefficients. Take note of the standard definitions below. \begin{itemize}
\item A polynomial $u \in \R[s]$ \emph{divides} $p \in \R[s]$, denoted by $u \mid p$, iff $p = uq$ for some $q \in \R[s]$, which can also be written as $q = p/u$. The expression $u \nmid p$ asserts there is no such $q$. 
\item A polynomial is \emph{monic} if its leading coefficient is $1$.
\item Let $u = \gcd(p,q)$ denote the unique monic greatest common divisor of $p$ and $q$. If $p \neq 0$, then $\gcd(p,0) = p/\mathrm{lc}(p)$, where $\mathrm{lc}(p) \in \R\setminus\{0\}$ is the leading coefficient of $p$. Also, $\gcd(0,0)$ is undefined. 
\item Polynomials $p,q$ are \emph{coprime} iff $\gcd(p,q)=1$. 
\item A polynomial $p \in \R[s]$ is \emph{reducible} if $p = uv$ for some $u,v \in \R[s]\setminus \R$. A polynomial is \emph{irreducible} iff it is both non-constant and not reducible \cite[Chapter 25]{vonzurgathenModernComputerAlgebra2013}.
\item A polynomial $p \in \R[s]$ is \emph{square-free}~\cite{gianniSquarefreeAlgorithmsPositive1996} over $\R$ iff $\forall g \in \R[s],\ (g^2 \mid p \implies g \in \R)$.
\item Let $p'$ denote the derivative of $p \in \R[s]$. \end{itemize}
\begin{proposition} \label{prop:mignotte}
	Let $p \in \R[s]$ be non-constant, and let $v:= \gcd(p,p')$. Then $q:=p/v \in \R[s]$ is square-free. Furthermore, $(u|p \implies u|q)$, for any irreducible $u \in \R[s]$. 
\end{proposition}
\begin{proof}
	Proposition 6.6 of \cite{mignotteMathematicsComputerAlgebra1992} states that $q$ is square-free.
	Now it follows from \cite[Corollary 1]{gianniSquarefreeAlgorithmsPositive1996} that $q = \prod_{i=1}^m f_i$, where $p = \prod_{i=1}^m f_i^{e_i}$ is the irreducible factorisation \cite[Chapter 3.4]{mignotteMathematicsComputerAlgebra1992} of $p$. If $u$ is an irreducible factor of $p$, then $u|f_i$ for some $i$, which implies $u|q$. 
\end{proof}
\begin{corollary} \label{cor:commonRoots}
	Let $p \in \R[s]$ be non-constant and $q:=p/\gcd(p,p')$. Then $p(s) = 0$ iff $q(s) = 0$. 
\end{corollary}
The polynomial $q = p/\gcd(p,p')$ is referred to as the \emph{square-free part} of $p$. Proposition \ref{prop:mignotte} demonstrates that $q$ contains every irreducible factor of $p$, but without multiplicities. 
\begin{proposition} \label{prop:repeatedRoots}
	Let $p,u \in \R[s]$ be non-zero. Then $u^2 \mid p$ iff both $u \mid p$ and $u \mid p'$.
\end{proposition}
\begin{proof}
	The result holds trivially if $u \in \R$. Assume $u$ is non-constant. 
	If $p = gu^2$ for some $g \in \R[s]$, then $p' = g'u^2 + 2gu$, and therefore $u \mid p'$. Suppose now that $p = bu$ for some $b \in \R[s]$. Then $p$ is non-constant, and \begin{equation} p' = b'u + bu' \neq 0 \label{eq:dp} \end{equation}
	Recall that $\deg(u') < \deg(u)$, and therefore $u \nmid u'$. If in addition $u \mid p'$, then \eqref{eq:dp} implies $b = cu$ for some $c \in \R[s]$, which in turn yields $p = cu^2$. 
\end{proof}
For any polynomial $p(s) = \sum_{i=0}^d p_i s^i \in \R[s]$, the inequalities $\|p\|_\infty \leq \|p\|_2 \leq d \|p\|_\infty$ hold, where $\|p\|_\infty:= \max \{ |p_i| \mid 0 \leq i \leq d\}$ and $\|p\|_2:= \sqrt{ \sum_{i=0}^d p_i^2} $ \cite[Chapter 3]{gerhardModularAlgorithmsSymbolic2005}. 
\section{Problem formulation} \label{sec:problem}
\subsection{Trajectories and observation maps}
\begin{definition}
	A \emph{trajectory} is a continuous function $x:[0,\infty) \to X$.
\end{definition}
A trajectory represents the evolution of a system over continuous time. We refer to $X$ as the \emph{state space} of the trajectory, which at this stage need only be a topological space. Note the variable $x$ is used to represent both state values and trajectories, but its type is always clear from the context. 

Given a finite set $\G$ of atomic propositions, an \emph{observation map} is a function $h:X \to 2^\G$, which associates each state with a set of propositions. Specifically, $h(x) \subset \G$ is viewed as the set of propositions satisfied by the state $x \in X$. Consequently, the set of states that satisfy the proposition $\g \in \G$ is given by
$ \llbracket \g \rrbracket := \{ x \in X \mid \g \in h(x) \}$. Any state space region that can be generated by taking a finite number of set unions, intersections and complements over the family $\llbracket \g \rrbracket, \g \in \G$ is called a \emph{region of interest} generated by $h$. 
\begin{remark} \label{rem:partition}
	Consider a map $h:X \to A$, with $A$ a finite set. The state space $X$ can be partitioned into the family of subsets $h^{-1}\{a\},a \in A$. By definition~\footnote{Recall that $h^{-1}B$ is the preimage of $B \subset A$ under $h$, as per Section \ref{sec:notation}. Here, $B = \{a\}$.}, $h$ is constant over each $h^{-1}\{a\}$. Moreover, if $h$ is an observation map, then $A = 2^\G$ and $h^{-1}\{a\} = \left( \bigcap_{\g \in a} \llbracket \g \rrbracket \right) \setminus \left( \bigcup_{\g \in \G \setminus a} \llbracket \g \rrbracket\right)$, which shows that every $h^{-1}\{a\}$ is a region of interest generated by $h$. 
\end{remark}
Clearly $h \circ x(t)$ is the set of atomic propositions satisfied by the trajectory $x:[0,\infty) \to X$ at time $t \geq 0$. Thus, the composition $h \circ x:[0,\infty) \to 2^\G$ is the primary object to be verified. Since it is a map from the non-negative reals into a finite set, we can expect $h \circ x$ to be piecewise constant when the trajectory and observation map are well-behaved. This property is typically expressed as a finite variability condition, which requires that $h \circ x$ change value only a finite number of times over any bounded time interval~\cite{alurBenefitsRelaxingPunctuality1996}.
\begin{definition}[Interval Sequence] \label{def:intervalSequence}
	An \emph{interval sequence} $ \mathcal{I} = \{I_k \subset [0,\infty) \mid k \in \N\} $ is a set of non-empty intervals satisfying
	\begin{enumerate}[label={\theenumi)}]
		\item  $\forall k \geq 0,\ \sup I_k = \inf I_{k+1} < \infty$ \label{cl:adjacent}
		\item $0 \in I_0$ and $\sup I_k \to \infty$ as $k \to \infty$ \label{cl:progress}
		\item $ \forall k \geq 0,\ I_k \cap I_{k+1} = \emptyset$ \label{cl:mutuallyDisjoint}
		\item $ \forall k \geq 0,\ \sup I_k \in I_k \cup I_{k+1}$. \label{cl:noGaps}
	\end{enumerate}
\end{definition}
Definition \ref{def:intervalSequence} implies $\bigcup \mathcal{I} = [0,\infty)$, and it is equivalent to \cite[Defintion 2.2.1]{alurBenefitsRelaxingPunctuality1996}, but has been stated here in a manner that is easier for us to check. Note it is impossible for both $I_k$ and $I_{k+1}$ to be singletons, as this would contradict \ref{cl:adjacent} together with \ref{cl:mutuallyDisjoint}. A function $z$ with domain $S$ is constant over $I \subset S$ iff $z(s_1) = z(s_2)$ for all $s_1,s_2 \in I$. By this definition, every function is constant over the empty set.   
\begin{definition}[Finite variability] \label{def:finiteVar}
	A function $z:S \to A$, with $S \subset [0,\infty)$ an interval, is of \emph{finite variability} iff there exists an interval sequence $\mathcal{I}$ such that $z$ is constant over $I \cap S$ for every $I \in \mathcal{I}$.
\end{definition}  
A trajectory $x$ is said to be of finite variability under an observation map $h$~\cite[Definition 3]{liuSynthesisReactiveSwitching2013} iff $h \circ x$ is of finite variability. Only pathological pairs of trajectories and observation maps fail to meet this criterion. We exclude such cases from consideration, because they are of no practical interest. 

\subsection{Temporal logic semantics for continuous-time signals}
In order to reason precisely about $h \circ x$, a temporal logic that can be interpreted over continuous-time signals is required. We adopt  MITL~\cite[Section 2]{alurBenefitsRelaxingPunctuality1996}, which can be interpreted over finite variability signals via \emph{timed state sequence} (TSS) representations. Note the definition below is a restatement of \cite[Definition 2.2.1]{alurBenefitsRelaxingPunctuality1996}. 
\begin{definition} \label{def:TSS}
	A TSS over a finite set $\G$ is a pair $(\zeta,\mathcal{I})$ such that $\mathcal{I}$ is an interval sequence and $\zeta:\N \to 2^\G$.
\end{definition}
This differs from the definition adopted in \cite{fainekosRobustnessTemporalLogic2009}, which replaces each interval in $\mathcal{I}$ with a single point sampled from it. 
Given a TSS $(\zeta,\mathcal{I})$ over $\G$, and an MITL formula $\varphi$ over $\G$, the satisfaction relation  $(\zeta,\mathcal{I}) \models \varphi$ is defined in \cite[Definition 2.3.2.1]{alurBenefitsRelaxingPunctuality1996}. A grammar that generates MITL formulae is presented in \cite[Definition 2.3.1.1]{alurBenefitsRelaxingPunctuality1996}. These precise definitions are not critical to the development of this paper so we do not repeat them here, but focus instead on presenting the pertinent relationships between our work and the wider literature. 
\begin{definition} \label{def:TSSrep}
	A TSS $(\zeta,\mathcal{I})$, with $\mathcal{I} = \{ I_k \subset [0,\infty) \mid k \in \N \}$, is a  \emph{TSS representation} of $z:[0,\infty) \to 2^\G$ iff
	$$ \forall k \geq 0,\ \forall t \in I_k,\ z(t):= \zeta(k). $$
\end{definition}

Clearly every TSS represents a unique continuous-time signal, which by Definition \ref{def:finiteVar} is of finite variability. Furthermore, given $z:[0,\infty) \to 2^\G$ of finite variability, it is clear how a TSS representation $(\zeta,\mathcal{I})$ may be constructed from the interval sequence in Definition \ref{def:finiteVar}. Finally by \cite[Remark 2.3.2.2]{alurBenefitsRelaxingPunctuality1996}, if $(\zeta_1, \mathcal{I}_1)$ and $(\zeta_2, \mathcal{I}_2)$ are both TSS representations of $z:[0,\infty) \to 2^\G$, then for any MITL formula $\varphi$, 
$$ (\zeta_2, \mathcal{I}_2) \models \varphi \iff  (\zeta_1, \mathcal{I}_1) \models \varphi.$$ 
MITL satisfaction can therefore be extended to continuous-time finite variability signals as follows.  
\begin{definition}[MITL satisfaction]
	Let $\G$ be a finite set of atomic propositions, $\varphi$ an MITL formula over $\G$, and $z:[0,\infty) \to 2^\G$ a function of finite variability. The relation $z \models \varphi$ holds iff there exists a TSS representation $(\zeta,\mathcal{I})$ of $z$ such that $ (\zeta,\mathcal{I}) \models \varphi$. Otherwise, $z \not\models \varphi$. 
\end{definition}
Having defined the satisfaction relation for well-behaved continuous-time signals precisely, we are now in a position to state the general problem. 
\begin{problem}[MITL Verification] \label{prob:general}
	Given an MITL formula $\varphi$ over a finite set of atomic propositions $\G$, and a trajectory $x:[0,\infty) \to X$ of finite variability under an observation map $h:X \to 2^\G$, determine whether
	$ h \circ x \models \varphi.$
\end{problem}
The works \cite{fainekosRobustnessTemporalLogic2009, malerMonitoringTemporalProperties2004} address this problem for various fragments of MITL. In this paper, we restrict attention to a fragment of MITL that has temporal operators only over $[0,\infty)$, which corresponds to LTL without the next operator. 
\subsection{LTL as a fragment of MITL}
The MITL operator $\mathscr{U}_I$ in \cite[Definition 2.3.2.1]{alurBenefitsRelaxingPunctuality1996} is a strict non-matching until operator~\cite{furiaExpressivenessMTLVariants2007}, which only imposes requirements at future times. The standard semantics of LTL~\cite{wolperConstructingAutomataTemporal2001} are stated in terms of the non-strict non-matching until $\until$, which includes the current time within its scope. The relationship between the two is formalized below~\footnote{Definition \ref{def:LTLnext} and Remark \ref{rem:otherUntils} have been modified to correct an error in the published version of this paper, D. Selvaratnam, M. Cantoni, J. M. Davoren, and I. Shames, ``Sampling polynomial trajectories for LTL verification,” \emph{Theoretical Computer Science}, vol. 897, pp. 135–163, Jan. 2022, \href{https://doi.org/10.1016/j.tcs.2021.10.024}{\texttt{doi: 10.1016/j.tcs.2021.10.024.}} The corrections appear in red.}.    
\begin{definition}[LTL without next] \label{def:LTLnext}
	An MITL formula $\varphi$, over a finite set of atomic propositions $\G$, is an $\LTL_\varobslash$ formula over $\G$ iff it is generated by the restricted grammar
	$$ \varphi::= \g \mid \lnot \varphi \mid \varphi \land \varphi \mid \varphi \until \varphi,$$
	where $\g \in \G$, and $\varphi_1 \until \varphi_2:= \varphi_2 \lor \Big(\varphi_1 \land \left(\textcolor{red}{(\varphi_1 \lor \varphi_2)} \mathscr{U}_{[0,\infty)} \varphi_2\right)\Big)$. 
\end{definition} 
In the context of trajectory verification, nothing is lost by omitting the LTL next operator from the above grammar. That operator would only have meaning for a fixed sampling strategy, which we do not wish to assume. In its place, nested $\until$ operators could be used to impose a desired sequence of transitions without specifying when they must occur.  
More generally, for $\LTL_\varobslash$ formulas, the precise timing information in $h \circ x$ is not pertinent to the verification. The MITL satisfaction question for a TSS then reduces to an LTL satisfaction question for a discrete word, as per the remark below. 
\begin{remark} \label{rem:TSS2LTL}
	If $\varphi$ is an $\LTL_\varobslash$ formula, then $(\zeta,\mathcal{I}) \models  \varphi$ holds for the TSS $(\zeta,\mathcal{I})$  iff $\zeta \models \varphi$ under standard discrete-time LTL semantics. 
\end{remark}
\begin{remark} \label{rem:otherUntils}
	Definition \ref{def:LTLnext} adopts the non-strict non-matching until $\until$ for consistency with the standard discrete-time semantics of LTL. \textcolor{red}{A non-strict matching until, $\varphi_1 \until^M \varphi_2:= \varphi_1 \land  \left(\varphi_1 \mathscr{U}_{[0,\infty)}(\varphi_1 \land \varphi_2)\right)$~\cite[Table 1]{furiaExpressivenessMTLVariants2007}, could just as well be adopted without affecting the results in this paper, as long as the corresponding semantics are adopted in discrete-time. When compared with \cite[Table 1]{furiaExpressivenessMTLVariants2007}, the expression for $\until$ in Definition \ref{def:LTLnext} has an extra disjunction in the first argument of $\mathscr{U}_{[0,\infty)}$. This disjunction is required for Remark \ref{rem:TSS2LTL} to hold. It takes care of edge cases that have $\varphi_1$ true on a right-closed interval, and $\varphi_2$ true on an adjacent left-open interval, as the discussion following \cite[Definition 4.2]{henzingerSymbolicModelChecking1994} explains in more detail; see also \cite[Sections 2.2 and 2.4.1]{schobbensAxiomsRealtimeLogics2002}. This subtlety arises for non-matching variants because strict inequality is a dense linear ordering on $\R$, but not on $\N$. For strict until variants, a similar difficulty arises near time 0, but is not so easily resolved by a translation of the formula. In this case, the continuous-time $(\zeta,\{ I_0,I_1, \hdots \}) \models \varphi$ cannot be inferred from the discrete-time $\zeta \models \varphi$, because a strict until does not constrain $\zeta(0)$ while the interval $I_0$ may include times greater than 0.} 
\end{remark}
Given LTL path checking has been well studied, and algorithmic solutions to it exist~\cite{markeyModelCheckingPath2003,basinOptimalProofsLinear2018}, the missing step in the $\LTL_\varobslash$ verification of $h \circ x$ is the construction of the first component $\zeta$ of its TSS representation, which we call an infinite trace.
\begin{definition}[Infinite trace] \label{def:inf_trace}
	A function $\zeta:\N \to 2^\G$ is an \emph{infinite trace} of $z:[0,\infty) \to 2^\G$ iff there exists an interval sequence $\mathcal{I}$ such that $(\zeta,\mathcal{I})$ is a TSS representation of $z$. 
\end{definition}
Thus we have shown that for $\LTL_\varobslash$ requirements, Problem \ref{prob:general} reduces to Problem \ref{prob:main} below. 
\begin{problem}[Infinite trace construction] \label{prob:main}
	Given a trajectory $x:[0,\infty) \to X$ of finite variability under an observation map $h:X \to 2^\G$, construct an infinite trace of $h \circ x$. 
\end{problem}

The remainder of this paper is devoted to Problem \ref{prob:main}. A complete algorithmic solution is provided for particular classes of trajectories and observation maps. 
\section{Sampling trajectories} \label{sec:samplingTheory}
\subsection{Trajectories and paths} \label{sec:paths}
A path is a standard object in topology, which can be used to describe the geometry and ordering of portions of a trajectory, while abstracting away the timing information.  
\begin{definition} 
	A \emph{path} is a continuous function $r:[0,1] \to X$. 
\end{definition} 
A path has a compact domain, which forces its image to also be compact. This need not be true of a trajectory, and so not every trajectory can be fully represented by a path. Consider however the path $r(s) = x(a(1-s) + bs)$, which describes the restriction of trajectory $x$ to $[a,b]$. The coordinate transformation $s \mapsto a(1-s) + bs$ preserves the geometry and direction of $x$. All the information relevant to the LTL verification of $h \circ x$ within the time interval $[a,b]$ is retained in $h \circ r$. 

The relationship between paths and trajectories in continuous time mirrors the relationship between finite and infinite words in discrete time. As shown below, the prefix of a trajectory is a path, and the suffix of a trajectory is a trajectory. 
\begin{definition}[Trajectory prefix and suffix] \label{def:trajPref}
	Let $x:[0,\infty) \to X$ be a trajectory. Then for a given $T \geq 0$, the path $x_{\leq T}:[0,1] \to X,\ x_{\leq T}(s) := x(Ts)$ is the \emph{prefix} of $x$, and the trajectory $x_{\geq T}:[0,\infty) \to X,\ x_{\geq T}(t) := x(t+T)$ is the \emph{suffix} of $x$.
\end{definition}
\begin{definition}[Path prefix and suffix] \label{def:pathPref}
	Let $r:[0,1] \to X$ be a path. Then for a given $T  \in [0,1]$, the path $r_{\leq T}:[0,1] \to X,\ r_{\leq T}(s) = r(Ts)$ is the \emph{prefix} of $r$, and the path $r_{\geq T}:[0,1] \to X,\ r_{\geq T}(s) := r(T + (1-T)s)$ is the \emph{suffix} of $r$.
\end{definition}
The concatenation operation defined below is called the path product in \cite{munkresTopology2000}. 
\begin{definition}[Path concatenation]
	Let $r_1, r_2:[0,1] \to X$ be paths such that $r_1(1) = r_2(0)$. Then the concatenation $r_1 \cdot r_2:[0,1] \to X$ is defined by
	$$r_1 \cdot r_2(s) :=  \begin{cases}
		r_1(2s),& \IF s \in [0,\frac{1}{2}],\\
		r_2(2s-1),& \IF s \in (\frac{1}{2},1]
	\end{cases}.$$
\end{definition}
The concatenation of two paths is a path \cite[Chapter 9]{munkresTopology2000}. 
Trajectories may evolve along the same path in different ways. These correspond to having $x = r \circ u$ for different mappings $u:[0,\infty) \to [0,1]$, which we refer to as \emph{motions}. 
\begin{definition} \label{def:directMotion}
	A continuous map $u:[0,\infty) \to [0,1]$ is a \emph{direct motion} iff it is non-decreasing, $u(0) = 0$, and $u(t) \to 1$ as $t \to \infty$. 
\end{definition}
If $u$ is a direct motion, then $r \circ u$ is called a \emph{direct trajectory}, which either arrives at the endpoint $r(1)$ in finite time, or approaches it asymptotically. Although many direct motions are possible, we can expect the resulting trajectories to retain similar LTL properties when defined along the same path. 
\begin{definition}
	A \emph{loop} is a path $r$ such that $r(1) = r(0)$. 
\end{definition}
The map $\mathsf{mod}:[0,\infty) \to [0,1)$, given by $\mathsf{mod}(s) := s - \lfloor s \rfloor$, returns $s$ modulo 1. 
\begin{definition}
	A mapping $u:[0,\infty) \to [0,1]$ is a \emph{cyclic motion} iff there exists a non-decreasing surjection $v:[0,\infty) \to [0,\infty)$ such that $u = \mathsf{mod} \circ v$. 
\end{definition} 
If $r$ is a loop, and $u$ is a cyclic motion, the trajectory $r \circ u$ moves perpetually around the loop in one direction. Such trajectories are referred to as \emph{cyclic}. They need not be periodic, because they can move around the loop at different speeds each time. Thus, with the aid of path concatenation and direct/cyclic motions, a rich class of trajectories can be constructed from a finite number of paths. 
\begin{remark}\label{rem:motionPlanning}
This framework of paths, motions and trajectories is consistent with the decomposition of the robotic motion planning problem into a \emph{path planning} stage followed by a \emph{path tracking} stage, discussed in \cite{verscheureTimeOptimalPathTracking2009}. Path planning generates the path $r$, and path tracking prescribes a motion $u$ along the path. Together, they yield the trajectory $x = r \circ u$, which constitutes the motion plan. 
\end{remark}

\subsection{Sampling functions for traces} \label{sec:samplingForTraces}
Let $(\zeta,\mathcal{I})$ be a TSS representation of $h \circ x$.
According to Definition \ref{def:TSSrep}, the infinite trace $\zeta$ is sampled from $h \circ x$ once within each interval of $\mathcal{I}$. A sampling function can be used to represent this, as done in \cite{fainekosRobustnessTemporalLogic2009}. 
\begin{definition}[Infinite sampling function] \label{def:infiniteSampling}
	A map $\sigma:\N \to [0,\infty)$ is an \emph{infinite sampling function} iff $\sigma$ is strictly increasing, $\sigma(0) = 0$ and $\sigma(k) \to \infty$ as $k \to \infty$.
\end{definition}
Observe that if $\sigma$ is an infinite sampling function, then $h \circ x \circ \sigma:\N \to 2^\G$ is a discrete word, but not necessarily an infinite trace (see Definition \ref{def:inf_trace}).  In the same way, paths can be sampled to obtain finite words. 
\begin{definition}[Finite sampling function] \label{def:finite_sampling}
	A map $\sigma:\{0,...,K\} \to [0,1]$ is a \emph{finite sampling function} iff $\sigma$ is strictly increasing, $\sigma(0) = 0$ and $\sigma(K) = 1$. 
\end{definition}
\begin{definition}[Sampling function] \label{def:sampling}
	A \emph{sampling function} is a finite sampling function or an infinite sampling function. 
\end{definition}
We refer to the values $\sigma(k)$ taken by a sampling function as \emph{checkpoints}. 
Beyond Definition \ref{def:infiniteSampling}, extra conditions on $\sigma$ are required to guarantee that $h \circ x \circ \sigma$ is an infinite trace. These are made explicit in the definition below, which generalises Definition \ref{def:inf_trace} to allow for finite traces.   
\begin{definition}[Trace] \label{def:trace}
	Let $A$ be a non-empty finite set, $\K \subset \N$ an interval containing $0$, and either $S = [0,1]$ or $S = [0,\infty)$. Then $\zeta:\K \to A$ is a \emph{trace} of $z:S \to A$ iff there exist both a sampling function $\sigma:\K \to S$ and set $S^*:=\{s^*_k \in S \mid k \in \K^\ominus \}$ such that all of the following hold:
	\begin{enumerate}[label={\theenumi)}]
		\item $\zeta = z \circ \sigma$ \label{cl:traceComposition}
		\item $\sup \sigma[\K] = \sup S \in \{1,\infty\}$ \label{cl:lengthOfTrace}
		\item $\forall k \in \K^\ominus,\ \sigma(k) \leq s^*_k \leq \sigma(k+1)$ \label{cl:betweenFrames}
		\item  $\forall k \in \K^\ominus,\ \forall s \in [\sigma(k), s^*_k),\ z(s) = z \circ \sigma(k)$ \label{cl:equalPre}
		\item  $\forall k \in \K^\ominus,\ \forall s \in (s^*_k, \sigma(k+1)],\ z(s) = z \circ \sigma(k+1)$ \label{cl:equalPost}
		\item $\forall k \in \K^\ominus,\ z(s^*_k) \in \{z \circ \sigma(k),\ z \circ \sigma(k+1) \}. $ \label{cl:inMid}
	\end{enumerate}
	The sampling function $\sigma$ is then said to \emph{record} a trace of $z$. 
	The checkpoints $\sigma(k)$ are the sample points, which are always distinct by Definitions \ref{def:infiniteSampling} -- \ref{def:sampling}. The set $S^*$ includes all times (or path parameter values) at which $z$ changes value, but $z$ is not required to change value at every $s^*_k \in S$. It is possible that $s^*_k = s^*_{k+1}$ for some $k$, and this case is discussed further in Remark \ref{rem:isolatedPoints}. 
	\begin{remark}[Finiteness of trace] \label{rem:traceFiniteness}
		Clause \ref{cl:lengthOfTrace} requires that $\sigma$ be a finite sampling function when $S = [0,1]$, and an infinite sampling function when $S = [0,\infty)$. Thus a trace of $h \circ y$ must have a finite domain if $y$ is a path, and domain $\N$ if $y$ is a trajectory. 
	\end{remark}
	\begin{remark}[Origins of trace definition]
		Definition \ref{def:trace} is based on \cite[Definition 2]{wongpiromsarnAutomataTheoryMeets2016}, which in turn derives from \cite{kloetzerFullyAutomatedFramework2008,liuSynthesisReactiveSwitching2013}. Note some differences that are introduced here for convenience. Firstly, Definition \ref{def:trace} defines a trace in terms of a sampling function, and generalises \cite[Definition 2]{wongpiromsarnAutomataTheoryMeets2016} to permit traces with finite domains. Secondly, \cite[Definition 2]{wongpiromsarnAutomataTheoryMeets2016} is stated for $x$ directly, whereas Definition \ref{def:trace} applies to $h \circ x$. Specifically, $\zeta:\N \to 2^\G$ is a trace of $h \circ x$ under Definition \ref{def:trace} iff it is a trace of $x$ under \cite[Definition 2]{wongpiromsarnAutomataTheoryMeets2016}.
	\end{remark}
	
\end{definition} 
The following result establishes that Definitions \ref{def:inf_trace} and \ref{def:trace} are consistent when $S = [0,\infty)$. Its proof is more involved than the simplicity of the result might suggest, because of the number of clauses in both Definitions \ref{def:trace} and \ref{def:intervalSequence} that must be verified.
\begin{proposition} \label{prop:consistentTrace}
	A mapping $\zeta:\N \to 2^\G$ is a trace of $z:[0,\infty) \to 2^\G$ iff it is an infinite trace of $z$.   
\end{proposition}
\begin{proof}
	$(\implies)$ Suppose $\zeta$ is a trace of $z$. Then by Remark \ref{rem:traceFiniteness}, there exist an infinite sampling function $\sigma:\N \to [0,\infty)$ and set $S^* := \{ s^*_k \geq 0 \mid k \in \N\}$ satisfying all clauses of Definition \ref{def:trace}. By Clause \ref{cl:traceComposition}, $\zeta = z \circ \sigma$. Since $\sigma$ is strictly increasing, Clause \ref{cl:betweenFrames} implies 
	$ 0 \leq s^*_0 \leq s^*_1 \hdots $, where each $s^*_k \in S^*$. 
	An interval sequence $\mathcal{I}$ is now constructed by choosing $s^*_{k-1},s^*_k \in S^*$ as the endpoints of every $I_k \in \mathcal{I}$. To ensure mutual disjointness, the right endpoint $s^*_k$ should be included in $I_k$ iff
	\begin{equation} \psi(k):\ z(s^*_k) = \zeta(k)  \text{ and }  s^*_k < s^*_{k+1}, \label{eq:predicate} \end{equation}
	where the predicate $\psi$, defined over all $k \in \N$, has been introduced to clarify the subsequent presentation.  
	Let
	\begin{equation} I_0 := \begin{cases} [0,s^*_0], & \IF \psi(0) \\
			[0,s^*_0), & \IF \lnot \psi(0) \end{cases},
		\label{eq:I0} \end{equation}
	and for each $k > 0$, \begin{equation} I_k:= \begin{cases}
			[s^*_{k-1}, s^*_k], & \IF \lnot \psi(k-1) \ \land \ \psi(k) \\
			(s^*_{k-1}, s^*_k], & \IF \psi(k-1) \ \land\ \ \psi(k) \\
			[s^*_{k-1}, s^*_k), & \IF \lnot \psi(k-1) \ \land \ \lnot \psi(k) \\
			(s^*_{k-1}, s^*_k), & \IF \psi(k-1) \ \land\  \lnot \psi(k)
		\end{cases}. \label{eq:I_k} \end{equation}
	Clearly the above case definition covers all possibilities. 
	It must be shown that $\mathcal{I}:= \{I_k \mid k \in \N \}$ is an interval sequence. Each $I_k$ is an interval by construction, so it must now be shown that they are non-empty. If $s^*_0 > 0$, then $I_0 \neq \emptyset$. Suppose on the other hand that $s^*_0 = 0$. Then because $\sigma$ is strictly increasing,
	$$0 = s_0 = \sigma(0) < \sigma(1) \leq s^*_1$$
	by Clause \ref{cl:betweenFrames}, which implies $\psi(0)$ is true. Equation \eqref{eq:I0} in turn implies $I_0 = [0,0]= \{0\}$. Consider now $k > 0$. If $s^*_{k-1} < s^*_k$, then $I_k \neq \emptyset$. Suppose instead that $s^*_{k-1} = s^*_k$. Then
	$$  \sigma(k-1) < s^*_{k-1} = \sigma(k) = s^*_k < \sigma(k+1) \leq s^*_{k+1},$$
	because $\sigma$ is strictly increasing. This implies $\psi(k)$ is true and $\psi(k-1)$ is false, and \eqref{eq:I_k} in turn implies $I_k = [s^*_{k-1},s^*_k] = \{s^*_k\}$. This establishes that $I_k \neq \emptyset$ for all $k \in \N$, as required by Definition \ref{def:intervalSequence}.  Given this, Clause \ref{cl:adjacent} clearly follows from \eqref{eq:I0} -- \eqref{eq:I_k}. To establish Clause \ref{cl:progress}, first note that $0 \in I_0$ by \eqref{eq:I0}. Now since $\sigma(k) \to \infty$ by Definition \ref{def:infiniteSampling}, Clause \ref{cl:betweenFrames} then implies $s^*_k = \sup I_k \to \infty$ as $k \to \infty$. Turning to the the next clause, Equations \eqref{eq:I0} -- \eqref{eq:I_k} imply $I_k \cap I_{k+1} \subset \{s^*_k \}$ for all $k \geq 0$. If $s^*_k \in I_k$, then \eqref{eq:I0} -- \eqref{eq:I_k} imply $\psi(k)$ is true, which further implies $s^*_k \notin I_{k+1}$. This establishes Clause \ref{cl:mutuallyDisjoint}. Conversely, if $s^*_k \notin I_k$, then $\psi(k)$ is false, which implies $s^*_k \in I_{k+1}$ by \eqref{eq:I_k}, establishing Clause \ref{cl:noGaps}. Thus $\mathcal{I}$ is an interval sequence. 
	
	It remains to be shown that $(\zeta,\mathcal{I})$ is a TSS representation of $z$. Choose any $s \in I_0 \subset [0,s^*_0]$. If $s \in [0,s^*_0)$, then Clause \ref{cl:equalPre} implies $z(s) = z \circ \sigma(0) =  \zeta(0)$. Suppose instead that $s = s^*_0$. Since $s \in I_0$, \eqref{eq:I0} then implies $\psi(0)$ is true, by which $z(s) = z(s^*_k) = \zeta(0)$. Thus $z(s) = \zeta(0)$ for all $s \in I_0$. Now choose any $k > 0$ and $s \in I_k \in [s^*_{k-1},s^*_k]$. If $s \in (s^*_{k-1},s^*_k) = (s^*_{k-1},  \sigma(k) ] \cup [\sigma(k), s^*_k)$, Clauses \ref{cl:equalPre} -- \ref{cl:equalPost} imply $z(s) =z \circ \sigma(k) = \zeta(k)$. Alternatively, if $s = s^*_k$, then $\psi(k)$ is true by \eqref{eq:I_k}, implying $z(s) = z(s^*_k) = \zeta(k)$. 
	The only remaining possibility for $s \in I_k$ is that $s = s^*_{k-1}$. In that case, $\psi(k-1)$ is false, which implies that $z(s^*_{k-1}) \neq \zeta(k-1)$ or $s^*_{k-1} = s^*_k$. If the latter, then $s = s^*_k$ and it has already been shown that $z(s) = \zeta(k)$. If the former, then Clause \ref{cl:inMid} implies $z(s) = z(s^*_{k-1}) = \zeta(k)$. It has been shown that, for any $k \geq 0$, for any $s \in I_k$, $z(s) = \zeta(k)$, satisfying Definition \ref{def:TSSrep}. Therefore $\zeta$ is an infinite trace of $z$ by Definition \ref{def:inf_trace}.  
	
	$(\impliedby)$ Suppose $(\zeta,\mathcal{I})$ is a TSS representation of $z$. A sampling function $\sigma$ and a set $S^*$ must now be constructed to satisfy all clauses of Definition \ref{def:trace}.
	Definition \ref{def:intervalSequence} implies each $I_k \in \mathcal{I}$ is non-empty. Define $S^*:= \{ s^*_k \geq 0 \mid k \in \N\}$ where $s^*_k:= \sup I_k$ for all $I_k \in \mathcal{I}$. Let $\sigma:\N \to [0,\infty)$ be a function such that $\sigma(0) = 0$ and $\sigma(k) \in I_k$ for all $k > 0$ (e.g., choose $\sigma(k)$ to be the midpoint of $I_k$). It then follows from Definition \ref{def:TSSrep} that $\zeta = z \circ \sigma$, satisfying Clause \ref{cl:traceComposition}.
	Note also that $0 \in I_0$ by Clause \ref{cl:progress}. Clause \ref{cl:adjacent} now implies that $\sigma$ is non-decreasing, and Clause \ref{cl:mutuallyDisjoint} then that it is strictly increasing. Clauses \ref{cl:adjacent} and \ref{cl:progress} also imply that $\inf I_{k+1} = \sup I_k \to \infty$, and therefore $\sigma(k+1) \to \infty$. Thus $\sigma$ is an infinite sampling function by Definition \ref{def:infiniteSampling}, which in particular, satisfies Clause \ref{cl:lengthOfTrace}. To establish Clause \ref{cl:betweenFrames}, $\sigma(k) \leq \sup I_k = s^*_k = \inf I_{k+1} \leq \sigma(k+1) $ for all $k \geq 0$. To establish Clause \ref{cl:equalPre}, choose any $k \geq 0$. If $\sigma(k) = s^*_k$, then Clause \ref{cl:equalPre} holds trivially for that $k$. Assume instead that $\sigma(k) < s^*_k$ and choose any $s \in [\sigma(k), s^*_k)$. Since $\sigma(k) \in I_k$ and $s^*_k = \sup I_k$, it follows that $s \in I_k$. Since also $\sigma(k) \in I_k$, Definition \ref{def:TSSrep} then implies $z(s) = z \circ \sigma(k) = \zeta(k)$. To establish Clause \ref{cl:equalPost}, choose any $k \geq 0$. If $ s^*_k = \sigma(k+1)$, then Clause \ref{cl:equalPost} holds trivially for that $k$. Assume instead that $s^*_k < \sigma(k+1) $ and choose any $s \in (s^*_k, \sigma(k+1)]$. Since $\sigma(k+1) \in I_{k+1}$ and $s^*_k = \sup I_k = \inf I_{k+1}$, it follows that $s \in I_{k+1}$. Since also $\sigma(k+1) \in I_{k+1}$, Definition \ref{def:TSSrep} then implies $z(s) = z \circ \sigma(k+1) = \zeta(k+1)$. Finally, Clause \ref{cl:inMid} is directly implied by Clause \ref{cl:noGaps} together with Definition \ref{def:TSSrep}.  
\end{proof}
\begin{remark} \label{rem:isolatedPoints}
	The proof of Proposition \ref{prop:consistentTrace} reveals that $S^*$ in Definition \ref{def:trace} consists of the endpoints of the intervals in the corresponding interval sequence in Definition \ref{def:inf_trace}. If some $I_m \in \mathcal{I}$ is a singleton, then $s^*_{m-1} = \sigma(m) = s^*_m$, which makes Clause \ref{cl:equalPre} vacuous at $k=m$ and Clause \ref{cl:equalPost} vacuous at $k = m-1$. In this case, it is possible that all three of the following hold:
	\begin{align} 
		& z(s) = z \circ \sigma(m-1) \text{ for all } s \in [\sigma(m-1), s^*_m), \label{eq:rem0} \\
		& z(s) = z \circ \sigma(m+1) \text{ for all } s \in (s^*_m, \sigma(m+1)], \\
		& z \circ \sigma(m-1) \neq z (s^*_m) \neq z \circ \sigma(m+1), \label{eq:rem1}
	\end{align}
	indicating that $z$ only holds its value at $s^*_m$ instantaneously. To record a trace, the sampling function must place checkpoint $\sigma(m)$ exactly at $s^*_m$. A uniform sampling strategy almost surely fails to achieve this. The coming sections develop a trace generation algorithm capable of computing $z(s^*_m) = h \circ x(s^*_m)$ exactly, for polynomial trajectories $x$, and observation maps $h$ that only generate semi-algebraic regions of interest.
\end{remark}

\subsection{Manipulating traces} \label{sec:traceOperations}
In this subsection it is shown how traces of paths may be combined, and traces of trajectories decomposed, into traces of paths. We first specify some relevant operations on discrete words. 
\begin{definition}[Word operations]
	The definitions below hold for any $\alpha:\K \to A$ and $\beta:\mathcal{L} \to A$, with $\mathcal{L} \subset \N$ an interval and $\K \subset \N$ a finite interval such that $0 \in \K \cap \mathcal{L}$. Define $\K \oplus \mathcal{L} := \{k + l \in \N \mid k \in \K,\ l \in \mathcal{L} \}$. 
	\begin{enumerate}[label={\theenumi)}]
		\item \emph{Letter repetition:} $a^\omega: \N \to \{a\}$, given $a \in A$.  
		\item \emph{Word repetition:} $\alpha^\omega:\N \to A$, with $\alpha^\omega(k):= \alpha( k \bmod|\K|)$ for all $k \in \N$.
		\item \emph{Word prefix:} $\beta_{<k} := \beta {\restriction_{\{ l \in \mathcal{L} \mid l < k \}}}$ and $\beta_{\leq k} := \beta{ \restriction_{\{ l \in \mathcal{L} \mid l \leq k \}}}$, given $k \in \N$. 
		\item \emph{Word concatenation:} $\alpha \cdot \beta:\K \oplus \mathcal{L} \to A$, where $$\alpha \cdot \beta(k):= \begin{cases}
			\alpha(k),& \IF k \in \K\\
			\beta(k - |\K|),& \IF k \notin \K
		\end{cases}.$$
		\item \emph{Letter concatenation:} $\alpha \cdot a := \alpha \cdot \vec{a}$, where $\vec{a}:\{0\} \to \{a\}$, given $a \in A$. 
	\end{enumerate}
\end{definition}
\begin{remark} \label{rem:ndsurjection}
	Every non-decreasing surjection $\gamma: [0,1] \to [0,1]$ is continuous, and can be viewed as a path reparameterization.
\end{remark}
\begin{proposition}[Path reparameterization] \label{prop:reparam}
	Let $r:[0,1] \to X$ be a path, $\gamma:[0,1] \to [0,1]$ a non-decreasing surjection, and $h:X \to A$, with $A$ finite. Every trace of $h \circ r$ is a trace of $h \circ r \circ \gamma$.  
\end{proposition}
\begin{proof}
	Let $\zeta$ be a trace of $h \circ r$. By Remark \ref{rem:traceFiniteness}, there exist a finite sampling function $\sigma:\{0,...,K\} \to [0,1]$ and set $S^* = \{ s^*_k \in [0,1] \mid 0 \leq k < K\}$ that satisfy all clauses of Definition \ref{def:trace}. Clause \ref{cl:traceComposition} implies $\zeta = h \circ r \circ \sigma$.
	First observe that, for any $s \in [0,1]$, $\gamma^{-1}\{s\} \subset [0,1]$ is non-empty because $\gamma$ is surjective, and closed because $\gamma$ is continuous (see Remark \ref{rem:ndsurjection}).
	Define $\theta: \{0,...,K\} \to [0,1]$ as
	\begin{equation} \theta(k):= \begin{cases}
			0,& \IF k = 0 \\
			\max \gamma^{-1} \{ \sigma(k) \},& \IF 0 < k \leq K
		\end{cases}, \label{eq:thetadef} \end{equation}
	and construct the set $U^* := \{ u^*_k \in [0,1] \mid 0 \leq k < K\}$ by choosing \begin{equation} u^*_k := \max \gamma^{-1} \{ s^*_k \} \label{eq:sdef} \end{equation} for all $0 \leq k < K$.
	Given that $\sigma$ and $S^*$ satisfy all clauses of Definition \ref{def:trace} for $z = h \circ r$, the task is now to show that $\theta$ is a sampling function, and that $\theta$ and $U^*$ satisfy all clauses of Definition \ref{def:trace} for $z = h \circ r \circ \gamma$.
	
	By Definition \ref{def:finite_sampling}, $\sigma(0) = \gamma(0) = 0$, implying that $\gamma \circ \theta(k) = \sigma(k)$ for all $0 \leq k \leq K$. It follows that $\gamma \circ \theta = \sigma$, and therefore $\zeta = h \circ r \circ \gamma \circ \theta$. Thus $\theta$ satisfies Clause \ref{cl:traceComposition}. Furthermore, $\sigma$ is injective by Definition \ref{def:finite_sampling}, which implies $ \gamma^{-1}\{\sigma(j)\} \cap \gamma^{-1}\{\sigma(k)\} = \emptyset$ whenever $j \neq k$. This in turn implies that $\theta$ is injective, and hence strictly increasing because $\gamma$ and $\sigma$ are both non-decreasing. Recall that $\gamma(1) = \sigma(K) = 1$, which implies that $\theta(K) = 1$. This establishes that $\theta$ is a finite sampling function, which in particular satisfies Clause \ref{cl:lengthOfTrace}. By Clause \ref{cl:betweenFrames},
	$ \gamma \circ \theta(k) \leq \gamma(u^*_k) \leq \gamma \circ \theta(k+1) $ for all $0 \leq k < K$.
	For $k = 0$, $\theta(0) = 0 \leq u^*_0$ by definition. Consider now any $0 < k < K$. 
	If $\gamma \circ \theta(k) < \gamma(u^*_k)$, then $\theta(k) < u^*_k$ because $\gamma$ is non-decreasing. Alternatively if $\gamma \circ \theta(k) = \gamma(u^*_k)$, then it follows from \eqref{eq:thetadef} and \eqref{eq:sdef} that $\theta(k) = u^*_k $. In the same way, it can be shown that $ u^*_k \leq \theta(k+1)$ for all $0 \leq k < K$, establishing Clause \ref{cl:betweenFrames}. 
	Turning to Clause \ref{cl:equalPre}, let $0 \leq k < K$ and suppose $\theta(k) \leq s < u^*_k$. Then
	$\sigma(k) = \gamma \circ \theta(k) \leq \gamma(s) \leq \gamma (u^*_k) = s^*_k$, which implies $h \circ r \circ \gamma(s) = h \circ r \circ \sigma(k) = h \circ r \circ \gamma \circ \theta(k).$ Clause \ref{cl:equalPost} can be established in a similar manner, and Clause \ref{cl:inMid} holds because $\sigma = \gamma \circ \theta$ and $s^*_k = \gamma(u^*_k)$ for all $0 \leq k < K$. 
\end{proof} 
\begin{corollary} \label{cor:prefixConversion}
	Let $r:[0,1] \to X$ be a path, $u:[0,\infty) \to [0,1]$ a direct motion, and $h:X \to A$, with $A$ finite. Then for any $T \geq 0$, every trace of $ h \circ r_{\leq u(T)}$ is a trace of $h \circ (r \circ u)_{\leq T}$.
\end{corollary}
\begin{proof}
	If $u(T) = 0$, then $u(t) = 0$ for all $t \in [0,T]$, and therefore $r_{\leq u(T)}(s) = r(0) =  (r \circ u)_{\leq T}(s)$ for all $s \in [0,1]$. If instead $u(T) > 0$, then $(r \circ u)_{\leq T} = r_{\leq u(T)} \circ \gamma$, where $ \gamma:[0,1] \to [0,1],\ \gamma(s):= \frac{u(Ts)}{u(T)}$ is a non-decreasing surjection.
\end{proof}
Propositions \ref{prop:snip} -- \ref{prop:cyclicTraj} below are consequences of the preceding developments, stated here without proof. 
\begin{proposition}[Snipping Lemma] \label{prop:snip}
	Let $r:[0,1] \to X$ be a path, and $h:X \to A$, with $A$ finite. Suppose $\sigma:\K \to [0,1]$ is a sampling function that records the trace $\zeta$ of $h \circ r$, and $T \in [\sigma(k),\sigma(k+1)]$ for some $k \in \K^\ominus$. If $h \circ r(T) = \zeta(k+1)$, then $\zeta_{\leq k+1}$ is a trace of $h \circ r_{\leq T}$. Otherwise, $\zeta_{\leq k} \cdot \zeta(k)$ is a trace of $h \circ r_{\leq T}$.
\end{proposition}
The concatenation of $\zeta_{\leq k}$ with $\zeta(k)$ in the Snipping Lemma is necessary to handle the case where $k = 0$, because Definition \ref{def:trace} requires the domain of a trace to contain at least two elements, but the domain of $\zeta_{\leq 0}$ is a singleton. 
\begin{proposition}[Path concatenation] \label{prop:concat}
	Let $r_1, r_2:[0,1] \to X$ be paths such that $r_1(1) = r_2(0)$, and $h:X \to A$, with $A$ finite. If $\zeta_1$ is a trace of $h \circ r_1$, and $\zeta_2$ a trace of $h \circ r_2$, then $\zeta_1 \cdot \zeta_2$ is a trace of $h \circ(r_1 \cdot r_2)$. 
\end{proposition}
We refer to a trajectory $x$ as \emph{invariant} under $h$ iff $h \circ x$ is constant.
\begin{proposition}[Invariant trajectories] \label{prop:invariantTraj}
	Let $x:[0,\infty) \to X$ be a trajectory, and $h:X \to A$, with $A$ finite. If $h \circ x$ is constant, then $(h \circ x(0))^\omega$ is a trace of $h \circ x$. 
\end{proposition}
\begin{proposition}[Trajectory suffixes and prefixes] \label{prop:trajPref}
	Let $x:[0,\infty) \to X$ be a trajectory, and $h:X \to A$, with $A$ finite. For any $T \geq 0$, if $\zeta_1$ is a trace of $h \circ x_{\leq T}$ and $\zeta_2$ a trace of $h \circ x_{\geq T}$, then $\zeta_1 \cdot \zeta_2$ is a trace of $h \circ x$. 
\end{proposition}

\begin{corollary}[Invariant suffixes] \label{cor:invariant}
	Let $x:[0,\infty) \to X$ be a trajectory, and $h:X \to A$, with $A$ finite.  Suppose there exists $T \geq 0$ such that $h \circ x_{\geq T}$ is constant. If $\zeta:\{0,...,K\} \to A$ is a trace of $h \circ x_{\leq T}$, then $\zeta \cdot \zeta(K)^\omega$ is a trace of $h \circ x$. 
\end{corollary}
\begin{proof} The result follows directly from Propositions \ref{prop:invariantTraj} and \ref{prop:trajPref}.\end{proof}
\begin{proposition}[Cyclic trajectories] \label{prop:cyclicTraj}
	Let $r:[0,1] \to X$ be a loop, $u:[0,\infty) \to [0,1]$ a cyclic motion, and $h:X \to A$, with $A$ finite. If $\zeta$ is a trace of $h \circ r$, then $\zeta^\omega$ is a trace of $h \circ r \circ u$. 
\end{proposition}

\begin{corollary}[Cyclic suffixes] \label{cor:cyclic}
	Let $x:[0,\infty) \to X$ be a trajectory, and $h:X \to A$, with $A$ finite. Suppose there exist a loop $r:[0,1] \to X$, cyclic motion $u:[0,\infty) \to [0,1]$, and $T \geq 0$ such that $x_{\geq T} = r \circ u$. If $\zeta_1$ is a trace of $h \circ x_{\leq T}$ and $\zeta_2$ a trace of $h \circ r$, then $\zeta_1 \cdot \zeta_2^\omega$ is a trace of $h \circ x$.  
\end{corollary}
\begin{proof} The result follows directly from Propositions \ref{prop:trajPref} and \ref{prop:cyclicTraj}.\end{proof}
Surprisingly, direct trajectories require a more verbose result. The complication arises when one asymptotically approaches a point on the boundary of a region of interest, because it never actually gets there. Notwithstanding this, all direct trajectories have invariant suffixes, as established below.
\begin{proposition}[Direct trajectories] \label{prop:direct}
	Let $r:[0,1] \to X$ be a path, $u:[0,\infty) \to [0,1]$ a direct motion, and $h:X \to A$, with $A$ finite. Suppose $\zeta:\{0,...,K\} \to A$ is a trace of $h \circ r$. Then, for $x := r \circ u$, there exists $T \geq 0$ such that $h \circ x_{\geq T}$ is constant. 
	Moreover, if at least one of the following hold:
	\begin{align}
		&\exists t \geq 0,\ u(t) =1, \label{eq:sub1} \\
		&\exists \varepsilon \in [0,1),\ \forall s \in (\varepsilon,1],\ h \circ r(s) = \zeta(K), \label{eq:endOnBoundary}
	\end{align}
	then $\zeta \cdot \zeta(K)^\omega$ is a trace of $h \circ x$. Otherwise, $\zeta_{<K} \cdot \zeta(K-1)^\omega$  is a trace of $h \circ x$.
\end{proposition}
\begin{proof}
	Since $\zeta$ is a trace of $h \circ r$, by Remark \ref{rem:traceFiniteness}, there exist a finite sampling function $\sigma:\{0,...,K\} \to [0,1]$ and set $S^* = \{ s^*_k \in [0,1] \mid 0 \leq k < K\}$ that satisfy all clauses of Definition \ref{def:trace}. Clause \ref{cl:traceComposition} implies $\zeta = h \circ r \circ \sigma$. There are now three cases to consider, which cover all possibilities. 
	\begin{itemize}
		\item Suppose \eqref{eq:sub1} holds, and choose any $T \geq 0$ such that $u(T) = 1$. Since $u$ is non-decreasing, it follows that $u(t) = 1$ for all $t \geq T$. Noting that $\sigma(K) = 1$, it follows that $h \circ r \circ u(t) = \zeta(K)$ for all $t \geq T$, implying that $h \circ (r \circ u)_{\geq T}$ is constant. Now Corollary \ref{cor:prefixConversion} implies that $\zeta$ is a trace of $h \circ (r \circ u)_{\leq T}$ because it is a trace of $h \circ r$, and applying Corollary \ref{cor:invariant} then yields the result. 
		
		\item \label{case:middle} Suppose \eqref{eq:endOnBoundary} holds. Choose $\varepsilon \in [0,1)$ as in \eqref{eq:endOnBoundary}. Since $u$ is non-decreasing and $u(t) \to 1$,
		\begin{equation} \exists T \geq 0,\ \forall t \geq T,\ \max\{\varepsilon,\sigma(K-1)\} < u(t) \leq 1 = \sigma(K), \label{eq:u_loc} \end{equation}
		by which \eqref{eq:endOnBoundary} implies
		$h \circ r \circ u(t) = \zeta(K)$ for all $t \geq T$. This shows that $h \circ (r \circ u)_{\geq T}$ is constant, and in particular that $h \circ r \circ u(T) = \zeta(K)$. Noting also that $u(T) \in (\sigma(K-1),\sigma(K)]$ by \eqref{eq:u_loc}, the Snipping Lemma then implies that $\zeta$ is a trace of $h \circ r_{\leq u(T)}$.  Corollary \ref{cor:prefixConversion} in turn implies that $\zeta$ is a trace of $h \circ (r \circ u)_{\leq T}$, and applying Corollary \ref{cor:invariant} then yields the result. 
		
		\item Suppose \eqref{eq:sub1} and \eqref{eq:endOnBoundary} are both false. If $s^*_{K-1}<1$, then Clause \ref{cl:equalPost} implies \eqref{eq:endOnBoundary}, which is a contradiction. Thus $s^*_{K-1}=1$. Since $u(t) \to 1$, it follows that
		\begin{equation} \exists T \geq 0,\ \forall t \geq T,\ \sigma(K-1) < u(t) < s^*_{K-1} = \sigma(K) = 1. \label{eq:u_s=1} \end{equation}
		Thus, $h\circ r \circ u(t) = \zeta(K-1)$ for all $t \geq T$. This shows that $h \circ (r \circ u)_{\geq T}$ is constant, and in particular that $h \circ r \circ u(T) = \zeta(K-1)$. Noting also that $u(T) \in (\sigma(K-1),\sigma(K))$ by \eqref{eq:u_s=1}, the Snipping Lemma then implies that $\zeta_{< K} \cdot \zeta(K-1)$ is a trace of $h \circ r_{\leq u(T)}$.  Corollary \ref{cor:prefixConversion} in turn implies that $\zeta_{< K} \cdot \zeta(K-1)$ is also a trace of $h \circ (r \circ u)_{\leq T}$, and applying Corollary \ref{cor:invariant} yields the result. 
	\end{itemize}
\end{proof} 
\begin{remark} \label{rem:practicalTrajectories}
	Only trajectories with cyclic or invariant suffixes are practically amenable to path checking, because in the absence of a model for the underlying dynamics, these are the only types of trajectories that offer information infinitely far into the future. Corollaries \ref{cor:invariant} and \ref{cor:cyclic} establish that both have lasso word traces constructed from the traces of at most two paths. LTL verification of these traces can be performed using the discrete path-checking algorithms of \cite{markeyModelCheckingPath2003,basinOptimalProofsLinear2018}.
\end{remark}
Remark \ref{rem:practicalTrajectories} implies that all practical solutions to Problem \ref{prob:main} can be constructed from solutions to the problem below, which is conducive to an algorithmic solution because it always has a finite output. Just as for trajectories, we say that a path $r$ is of finite variability under an observation map $h$ iff $h \circ r$ is of finite variability. 
\begin{problem}[Finite trace construction] \label{prob:practical}
	Given a path $r:[0,1] \to X$ of finite variability under an observation map $h:X \to 2^\G$, construct a trace of $h \circ r$. 
\end{problem}
Observe that this places no restrictions on the LTL formula that can be satisfied. For any LTL formula $\varphi$, there exists a non-deterministic Buchi automaton (NBA) that accepts all and only the words satisfying $\varphi$ \cite[Corollary 23]{vardiAutomatatheoreticApproachLinear1996}. Furthermore, if the language accepted by an NBA is nonempty, it contains a lasso word \cite[Proposition 3.3.1]{khoussainovAutomataTheoryIts2001}. The remark below is a corollary of this. 
\begin{remark} \label{rem:lasso}
	An LTL formula is satisfiable iff there exists a lasso word that satisfies it. 
\end{remark}
\begin{remark}[Implications for robotic motion planning] \label{rem:motionPlanningImplications}
In Corollary \ref{cor:cyclic} and Propositions \ref{prop:cyclicTraj} -- \ref{prop:direct}, the trace of $h \circ r \circ u$ depends primarily on the path $r$, and is invariant to the motion $u$ under only mild conditions. In the context of robotic motion planning (see Remark \ref{rem:motionPlanning}), this means that verification of the entire motion plan depends only on the path $r$, provided that the path tracking stage is restricted to produce a suitable class of motions. 
\end{remark}
\subsection{Topological conditions on sampling functions} \label{sec:abstract}
The result below provides a topological condition that is sufficient for $h \circ r$ or $h \circ x$ to be constant over an interval. 
\begin{corollary} \label{cor:constantZ}
	Let $z:S \to A$, with $A$ finite and $S \subset \R$ an interval, and let $I \subset S$ also be an interval. If
	\begin{equation} 
		\forall a \in A,\ I \cap \Bd(z^{-1}\{a\}|S) = \emptyset, \label{eq:noInter} 
	\end{equation}
	then $z$ is constant over $I$. 
\end{corollary}
\begin{proof}
	The result is trivial for $I = \emptyset$, so assume $I \neq \emptyset$. For all $a \in A$, $z^{-1}\{a\} \subset S$. In particular, choose $a = z(s^\star)$ for some $s^\star \in I$. Clearly $ s^\star \in I \cap z^{-1}\{a\}$, and if \eqref{eq:noInter} holds, then Lemma \ref{lem:noCrossing} implies $I \subset z^{-1}\{a\}$. Thus, $z[I] \subset \{a\}$.  
\end{proof}
A necessary topological condition for the situation described in Remark \ref{rem:isolatedPoints} is given in the next result. Specifically, that situation may only occur at isolated points. 
\begin{lemma} \label{lem:isolatedZ}
	Let $z:S \to A$, with $A$ finite and $S \subset \R$ an interval. Let $s_1, s^\star, s_2 \in S$, where $s_1 \leq s^\star \leq s_2$, and suppose $z$ is constant on $[s_1,s^\star)$ and on $(s^\star,s_2]$. If $z(s_1) \neq a \neq z(s_2)$, where $a := z(s^\star)$, then $s^\star \in \Iso(z^{-1} \{ a \})$. 
\end{lemma}
\begin{proof}
	Suppose $z(s_1) \neq a = z(s^\star) \neq z(s_2)$. Then clearly $s_1 < s^\star < s_2$. Choose $c_1 \in (s_1,s^\star)$ and $c_2 \in (s^\star, s_2)$. By assumption, $z(s) = z(s_1) \neq a$ for all $s_1 \leq s < s^\star$, and $z(s) = z(s_2) \neq a$ for all $s^\star < s \leq s_2$. Thus $(c_1,c_2) \cap z^{-1}\{a\} = \{s^\star\}$, which implies $s^\star \in \Iso(z^{-1}\{a\})$.
\end{proof}
We now combine these results to derive sufficient conditions for a sampling functions to record a trace. Though abstract, the result is crucial to the construction of our algorithm in Section \ref{sec:numerical}. Although the focus of subsequent sections is on paths, notice that the result is also applicable to trajectories.  
\begin{theorem} \label{thm:topological}
	Suppose that either $S = [0,1]$ and $\K \subset \N$ is finite, or $S = [0,\infty)$ and $\K = \N$. 
	Let $r:S \to X$ be continuous, and $h:X \to A$, with $A$ finite. 
	Define $R_a:= r^{-1}  h^{-1} \{a\}$ for $a \in A$, and
	\begin{equation} \Gamma := \bigcup_{a \in A} \Bd \left( R_a \mid S \right), \quad \Theta :=  \bigcup_{a \in A} \Iso( R_a ). \label{eq:GammaTheta} \end{equation}
	If $\sigma:\K \to S$ is a sampling function such that
	\begin{enumerate}[label = {\roman*)}]
		\item  $\Theta \subset \sigma[\K]$, \label{cond:theta}
		\item $\forall k \in \K^\ominus,\ \big|\Gamma \cap [\sigma(k), \sigma(k+1)]\big| \leq 1$, \label{cond:gamma}
	\end{enumerate}
	then $h \circ r \circ \sigma $ is a trace of $h \circ r$.
\end{theorem}
Before proceeding with the proof, we pause to interpret this theorem. Each $H_a:= h^{-1}\{a\} \subset X$ is a state space region over which $h$ is constant, and the family $H_a,a\in A$ partitions $X$ (see Remark \ref{rem:partition}). Given the path (or trajectory) $r$, $R_a \subset S$ is the set of parameter values $s$ at which $r(s) \in H_a$. The set $\Gamma \subset S$ contains only the parameter values at which $r$ enters or leaves some $H_a$, and $\Theta \subset S$ contains the parameter values at which the the path locally touches some $H_a$ at only a single point. Condition \ref{cond:theta} forces a checkpoint at such isolated points, and Condition \ref{cond:gamma} ensures there can be no more than one boundary crossing between consecutive checkpoints. Since $S$ is an interval and $R_a \subset S$, it follows that $\Theta \subset \Gamma$ by Proposition \ref{prop:isoInBd}. 
\begin{proof}
	Suppose Conditions \ref{cond:theta} and \ref{cond:gamma} hold. Clauses \ref{cl:traceComposition} and \ref{cl:lengthOfTrace} of Definition \ref{def:trace} follow directly from the theorem hypotheses. Below, we construct the set $S^* := \{s^*_k \in S \mid k \in \K^\ominus\}$, which is then shown to satisfy the remaining clauses of Definition \ref{def:trace}. 
	For all $k \in \K^\ominus$, define
	\begin{equation} s^*_k := \begin{cases}
			s,& \IF \Gamma \cap [\sigma(k), \sigma(k+1)] = \{ s \} \\
			\sigma(k),& \IF \Gamma \cap [\sigma(k), \sigma(k+1)] = \emptyset
		\end{cases}. \label{eq:t} \end{equation}
	Condition \ref{cond:gamma} ensures the above case definition is exhaustive. 
	Clearly $\sigma(k) \leq s^*_k \leq \sigma(k+1)$ for all $k \in \K^\ominus$, satisfying Clause \ref{cl:betweenFrames}.
	
	Suppose, for a contradiction, that there exists $k \in \K^\ominus$ for which $[\sigma(k),s^*_k) \cap \Gamma \neq \emptyset$. Then there exists $s \in \Gamma$ such that
	\begin{equation} \sigma(k) \leq s < s^*_k. \label{eq:upperInterval}\end{equation}
	Since $s \in [\sigma(k),s^*_k) \subset [\sigma(k), \sigma(k+1)]$ and $s \in \Gamma$, Condition \ref{cond:gamma} implies $\Gamma \cap [\sigma(k), \sigma(k+1)] = \{ s \}$.  Then $s=s^*_k$ by \eqref{eq:t}, which contradicts \eqref{eq:upperInterval}. Thus, $[\sigma(k),s^*_k) \cap \Gamma = \emptyset$ for all $k \in \K^\ominus$.  Further, using \eqref{eq:GammaTheta}, $$ \forall k \in \K^\ominus,\ \forall a \in A,\ [\sigma(k),s^*_k) \cap \Bd(R_a|S) = \emptyset. $$ Corollary \ref{cor:constantZ} then implies $h \circ r$ is constant over $[\sigma(k),s^*_k) $, whereby
	\begin{equation} \forall k \in \K^\ominus,\ \forall s \in [\sigma(k), s^*_k),\ h \circ r (s) = h \circ r \circ \sigma(k), \label{eq:prev} \end{equation}
	which establishes Clause \ref{cl:equalPre}.
	
	Suppose now there exists $k \in \K^\ominus$ for which $(s^*_k,\sigma(k+1)] \cap \Gamma \neq \emptyset$. Then there exists $s \in \Gamma$ such that
	\begin{equation} s^*_k < s \leq \sigma(k+1). \label{eq:lowerInterval}\end{equation}
	Since $s \in (s^*_k,\sigma(k+1)] \subset [\sigma(k), \sigma(k+1)]$ and $s \in \Gamma$, Condition \ref{cond:gamma} implies $\Gamma \cap [\sigma(k), \sigma(k+1)] = \{ s \}$.  Then $s=s^*_k$ by \eqref{eq:t}, which contradicts \eqref{eq:lowerInterval}. Thus, $(s^*_k,\sigma(k+1)] \cap \Gamma = \emptyset$ for all $k \in \K^\ominus$.  Moreover, using \eqref{eq:GammaTheta}, $$ \forall k \in \K^\ominus,\ \forall a \in A,\ (s^*_k,\sigma(k+1)] \cap \Bd(R_a|S) = \emptyset.$$ Corollary \ref{cor:constantZ} then implies $h \circ r$ is constant over $(s^*_k,\sigma(k+1)] $, therefore
	\begin{equation} \forall k \in \K^\ominus,\ \forall s \in (s^*_k,\sigma(k+1)],\ h \circ r (s) = h \circ r \circ \sigma(k+1), \label{eq:post} \end{equation}
	which establishes Clause \ref{cl:equalPost}.
	
	Now let $k \in \K^\ominus$, and suppose that both the following hold:
	\begin{align}
		h \circ r \circ \sigma(k) &\neq h \circ r(s^*_k), \label{eq:notPrev}\\
		h \circ r(s^*_k) &\neq h \circ r \circ \sigma(k+1). \label{eq:notPost}
	\end{align}
	Together with \eqref{eq:prev} and \eqref{eq:post}, Lemma \ref{lem:isolatedZ} then implies $s^*_k \in \Iso\left( R_a \mid S \right) \subset \Theta$, where $a = h \circ r (s^*_k)$. Condition \ref{cond:theta} further implies $s^*_k = \sigma(j)$ for some $j \in \K$, and \eqref{eq:t} in turn implies $\sigma(k) \leq \sigma(j) \leq \sigma(k+1)$. 
	However \eqref{eq:notPrev} and \eqref{eq:notPost} imply $j \notin \{k,k+1\}$. Since there is no natural number strictly between $k$ and $k+1$, a contradiction ensues. 
	Therefore, for any $k \in \K^\ominus$, \eqref{eq:notPrev} and \eqref{eq:notPost} cannot both be true, and Clause \ref{cl:inMid} has been established. 
\end{proof}
\begin{remark} \label{rem:isoTheta}
	The final paragraph of the proof demonstrates that if \eqref{eq:rem0} -- \eqref{eq:rem1} of Remark \ref{rem:isolatedPoints} hold for $z = h \circ r$, then $s^*_m \in \Theta$. \end{remark}
We conclude this section with a lemma that simplifies the application of Theorem \ref{thm:topological} to an observation map, which has codomain $A = 2^\G$. The result introduces a computationally cheaper over-approximation $\hat{\Gamma}$ to the set of boundary crossing points $\Gamma$.  
\begin{lemma} \label{lem:point2subset}
	Let $S \subset \R$ be an interval, $r:S \to X$ a continuous function, and $h:X \to 2^\G$, with $\G$ finite. Let $ \llbracket \g \rrbracket := \{ x \in X \mid \g \in h(x) \}$ for $\g \in \G$, and $R_a := r^{-1} h^{-1}\{a\} $ for $a \subset \G$. Then 
	\begin{align}
		\Gamma:=  \bigcup_{a \in  2^\G} \Bd (R_a \mid S)  \subset \bigcup_{\g \in \G}  \Bd \left(r^{-1} \llbracket \g \rrbracket \mid S \right) =: \hat{\Gamma}. \label{eq:bd}
	\end{align}
\end{lemma}
\begin{proof}
	Let $\llbracket \g \rrbracket^c := X \setminus \llbracket \g \rrbracket$. For any $a \subset \G$, 
	$$h^{-1}\{a\} = \left( \bigcap_{\g \in a} \llbracket \g \rrbracket \right) \setminus \left( \bigcup_{\g \in \G \setminus a} \llbracket \g \rrbracket\right) = \left( \bigcap_{\g \in a} \llbracket \g \rrbracket \right) \cap \left( \bigcap_{\g \in \G \setminus a} \llbracket \g \rrbracket^c\right),$$
	and therefore
	\begin{equation} R_a = \left( \bigcap_{\g \in a} r^{-1} \llbracket \g \rrbracket  \right) \cap \left( \bigcap_{\g \in \G \setminus a} \left( r^{-1} \llbracket \g \rrbracket \right)^c\right), \label{eq:preimageHa} \end{equation}
	where $r^{-1}( \cdot )^c := S \setminus r^{-1}( \cdot )$. 
	Applying Proposition \ref{prop:BdIntersection}, and noting that the boundary of a set is equal to the boundary of its complement,
	\begin{align*} 
		\Bd (R_a \mid S )&  \subset \left( \bigcup_{\g \in a} \Bd( r^{-1} \llbracket \g \rrbracket \mid S) \right) \cup \left( \bigcup_{\g \in \G \setminus a} \Bd \left((r^{-1} \llbracket \g \rrbracket)^c \mid S \right) \right)\\
		& = \bigcup_{\g \in \G} \Bd( r^{-1} \llbracket \g \rrbracket \mid S).
	\end{align*}
	Since the above hold for any $a \subset \G$, \eqref{eq:bd} then follows. 
\end{proof}
The set $\hat{\Gamma}$ in \eqref{eq:bd} is a union of $|\G|$-many sets, a saving when compared with $\Gamma$, which is a union of $2^{|\G|}$-many sets. The set inclusion in Lemma \ref{lem:point2subset} permits us to replace $\Gamma$ with $\hat{\Gamma}$ in Condition \ref{cond:gamma} of Theorem \ref{thm:topological}, without otherwise changing the result. 
\section{Sampling polynomial paths} \label{sec:concrete}
\subsection{Polynomial path and region assumptions}
In this section, we focus on solving a special case of Problem \ref{prob:practical} in which $r$ is a polynomial and $h$ only generates semi-algebraic regions of interest. Proposition \ref{prop:concat} on path concatenation allows the results of this section to be directly extended to polynomial splines. 

\begin{assumption}[Polynomial paths and region boundaries] \label{ass:main}
	Let $M, n \geq 1$. 
	\begin{enumerate}[label = {\roman*)}]
		\item \label{cl:h} Given $\G := \{ \g_1,...,\g_M\}$, let 
		\begin{equation}h:\R^n \to 2^\G,\  h(x):= \{ \g_i \in \G \mid g_i(x) \leq 0\}, \label{eq:h} \end{equation}
		where $g_i:\R^n \to \R$ is continuously differentiable for each $1 \leq i \leq M$. 
		\item \label{cl:gi} Suppose in addition that $g_i \in \R[x_1,\hdots,x_n]$ for each $1 \leq i \leq M$. 
		\item \label{cl:r} Let $r = \rho {\restriction_{[0,1]}}$ for some $\rho \in \R[s]^n$. 
	\end{enumerate}
\end{assumption}
Thus, $M \geq 1$ is the number of atomic proposition and $n \geq 1$ is the dimension of the state space. Under Clause \ref{cl:gi}, $h$ can be made to generate any semi-algebraic region of interest by choosing $M$ and the $g_i$ appropriately.
Note that $\rho:\R \to \R^n$ in Clause \ref{cl:r} is a vector function, each component of which is a univariate polynomial, and thus $r:[0,1] \to \R^n$. The key property that makes an algorithmic solution possible is that every $g_i \circ \rho$ is also a univariate polynomial, having a finite number of roots if non-zero. Lemma \ref{lem:IsoPoints} and Corollary \ref{cor:3cond} in the next subsection characterize the isolated points discussed in Remarks \ref{rem:isolatedPoints} and \ref{rem:isoTheta}. They provide insight in a more general context that only requires the $g_i$ to be continuously differentiable, which is the reason for separating Clauses \ref{cl:h} and \ref{cl:gi} in Assumption \ref{ass:main}. 

Take note of the relationship 
\begin{equation} \forall \g_i \in \G,\  r^{-1} \llbracket \g_i \rrbracket = (g_i \circ r)^{-1}(-\infty, 0]. \label{eq:giRelship} \end{equation}
The following lemma assures us that if $s \in [0,1]$ lies on the relative boundary of some $ r^{-1} \llbracket \g_i \rrbracket$, then for that $i$, we have $g_i \circ r(s) = 0$. 
\begin{lemma} \label{lem:zeroOnBoundary}
	Let $g:S \to \R$ be continuous, with $S \subset \R$ an interval. Then
	\begin{equation*} \Bd\left(g^{-1} (- \infty, 0] \ \big| \ S\right) \subset g^{-1}\{0\}.\end{equation*} 
\end{lemma}
\begin{proof}
	Let $G:=g^{-1} (- \infty, 0]$ and $s^* \in S$. Suppose that $g(s^*) > 0$. Since $g$ is continuous, there exists some $U$ that is an $S$-neighbourhood of $s^*$, such that $g(s)>0$ for all $s \in U$. Clearly $U \cap G = \emptyset$, and therefore $s^*$ is not a boundary point of $G$ relative to $S$. Similarly, if $g(s^*) < 0$, there exists some $U$ that is an $S$-neighbourhood of $s^*$, such that $g(s)<0$ for all $s \in U$. In this case $U \subset G$, and therefore $s^*$ is not a boundary point of $G$ relative to $S$. This proves that 
	$g(s^*) \neq 0 \implies s^* \notin \Bd(G|S) $.
\end{proof}
The next result implies that if $g_i \circ r$ is the zero polynomial, or has no roots in $[0,1]$, then the relative boundary of $r^{-1} \llbracket \g_i \rrbracket \subset [0,1]$ is empty. 
\begin{corollary} \label{cor:zeroPoly}
	Let $g:S \to \R$ be continuous, with $S \subset \R$ an interval. If $\Bd\big(g^{-1}(-\infty,0] \mid S \big) \neq \emptyset,$ then $\{0\} \subsetneq g[S]$. 
\end{corollary}
\begin{proof}
	Suppose $\Bd\big(g^{-1}(-\infty,0] \mid S \big) \neq \emptyset$. Then Lemma \ref{lem:zeroOnBoundary} implies $\{0\} \subset g[S]$.  If $\{0\} = g[S]$, then
	$ \Bd\big(g^{-1}(-\infty,0] \mid S \big) = \Bd\big(S \mid S \big) = \emptyset$ by Proposition \ref{prop:bdX}, which is a contradiction. 
\end{proof}
We now pause to consider polynomial trajectories, which (if non-zero) have unbounded images. This case violates Clause \ref{cl:r} of Assumption \ref{ass:main}, however the result below reveals that such  trajectories always have invariant suffixes when the remaining clauses hold. Therefore, according to Corollary \ref{cor:invariant}, the trace of a polynomial trajectory can still be constructed from the trace of a suitable prefix, which does satisfy Assumption \ref{ass:main}. Hence, traces of polynomial trajectories can also be computed within the current framework. 
\begin{proposition}[Unbounded polynomial trajectories] \label{prop:unboundedPoly}
	Let Clauses \ref{cl:h} and \ref{cl:gi} of Assumption \ref{ass:main} hold, $\rho \in \R[t]^n$, $x = \rho {\restriction_{[0,\infty)}}$, and $L \geq 0$. Define
	\begin{align*} 
		\mathcal{C}& :=\left\{ i \in \{1,...,M\} \mid \{0\} \subsetneq g_i \circ x \big[[0,\infty)\big]  \right\}, \\
		P & := \prod_{i \in \mathcal{C}} g_i \circ \rho.
	\end{align*}
	If $P(s) \neq 0$ for all $s \geq L$, then $ h \circ x_{\geq L}$ is constant.
\end{proposition}
\begin{proof}
	Suppose $h \circ x_{\geq L}$ is non-constant. Then by Corollary \ref{cor:constantZ}, 
	$$ \exists a \in 2^\G,\ [L,\infty) \cap \Bd\big( (h \circ x)^{-1} \{a\} \mid [0,\infty) \big) \neq \emptyset.$$
	Noting that $(h \circ x)^{-1} \{a\} = x^{-1} h^{-1} \{a\} =:R_a$ for $a \subset \G$, this implies
	$$ [L,\infty) \cap \bigcup_{a \in 2^\G} \Bd \left( R_a \mid [0,\infty) \right) \neq \emptyset,$$
	and Lemma \ref{lem:point2subset} in turn implies
	$$ [L,\infty) \cap \bigcup_{\g \in \G}  \Bd \left(x^{-1} \llbracket \g \rrbracket \mid [0,\infty) \right) \neq \emptyset .$$
	Recalling that $x^{-1} \llbracket \g_i \rrbracket = (g_i \circ x)^{-1}(-\infty,0]$ for every $\g_i \in \G$,
	$$ [L,\infty) \cap \bigcup_{i=1}^M \Bd \left( (g_i \circ x)^{-1}(-\infty,0] \mid [0,\infty) \right) \neq \emptyset.$$
	Lemma \ref{lem:zeroOnBoundary} and Corollary \ref{cor:zeroPoly} then imply
	$$ \exists s \geq L,\ \exists i \in \mathcal{C},\ g_i \circ x(s) = 0. $$
	Thus $P(s) = 0$ for some $s \geq L$, which establishes the contrapositive of the result. 
\end{proof}
\begin{remark}
	To obtain an invariant $x_{\geq L}$ in Proposition \ref{prop:unboundedPoly}, $L$ can be any strict upper bound on the real roots of $P$. Various upper bounds for polynomial roots are provided in \cite[Theorem 4.2]{mignotteMathematicsComputerAlgebra1992}. 
\end{remark}
\subsection{Isolated point characterisation} \label{sec:isolatedPointCharacterisation}
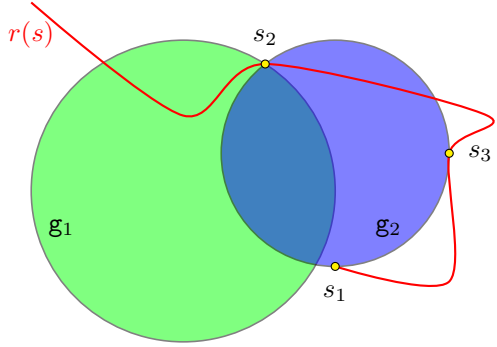
\begin{figure}
	\centering
	\begin{tikzpicture}
		\tkzDefPoint(0,0){A}  
		\tkzDefPoint(2,0.5){B}  
		\tkzDefPoint(2,-1){C}  
		\tkzDefPoint(3.5,0.5){M2}  
		\tkzInterCC[R](A,2 cm)(B,1.5 cm) \tkzGetPoints{M1}{N1}
		\tkzDrawCircle[R, fill=green, opacity=0.5](A,2.cm) 
		\tkzDrawCircle[R, fill=blue, opacity=0.5](B,1.5cm) 
		\draw[red, thick] plot [smooth] coordinates {(-2,2.5) (0,1) (M1) (4,1) (3.5,0.5) (3.5,-1.2) (C)}; 
		\tkzDrawPoints[color=black, fill=yellow](M1,M2,C)
		\node[label=above:$s_2$] at (M1) {};
		\node[label=right:$s_3$] at (M2) {};
		\node[label=below:$s_1$] at (C) {};
		\node[label=below:\textcolor{red}{$r(s)$}] at (-2,2.5) {};
		\node at (-1.6,-0.5) {$\g_1$};
		\node at (2.7,-0.5) {$\g_2$};
	\end{tikzpicture}
	\caption{Points $s_1, s_2, s_3 \in [0,1]$ depict the three types of isolated points in \eqref{eq:onBoundary} -- \eqref{eq:bounce}, respectively, from each of which follow \ref{pt:onBoundary} -- \ref{pt:bounce} of Corollary \ref{cor:3cond}, respectively. }
	\label{fig:isoPoints}
\end{figure}
The next result characterises the isolated points in $\Theta$ in terms of the relative boundaries of the $r^{-1} \llbracket \g_i \rrbracket$, and the derivatives $(g_i \circ r)'$. A more intuitive restatement is then provided in Corollary \ref{cor:3cond}, along with an illustration in Figure \ref{fig:isoPoints}. 
\begin{lemma} \label{lem:IsoPoints}
	Suppose Clause \ref{cl:h} of Assumption \ref{ass:main} holds. Let $r:[0,1] \to \R^n$ also be continuously differentiable, and $\llbracket \g_i \rrbracket := \{ x \in \R^n \mid g_i(x) \leq 0 \}$.
	If
	$$ s \in \Theta:=\bigcup_{a \in 2^\G} \Iso(R_a) ,$$ 
	where $R_a := r^{-1}h^{-1}\{a\}$ for $a \subset \G$, then at least one of the following hold:
	\begin{align}
		\exists i \in \{1,...,M\},\ & s \in \Bd( r^{-1} \llbracket \g_i \rrbracket  \mid [0,1]) \cap \{0,1\} \label{eq:onBoundary} \\
		\exists i,j \in \{1,...,M\},\ j\neq i,\ & s \in \Bd( r^{-1} \llbracket \g_i \rrbracket  \mid [0,1]) \cap \Bd( r^{-1}\llbracket \g_j \rrbracket \mid [0,1]) \label{eq:doubleCrossing} \\
		\exists i \in \{1,...,M\},\ & s \in \Bd( r^{-1} \llbracket \g_i \rrbracket  \mid [0,1]) \land (g_i \circ r)'(s) = 0. \label{eq:bounce}
	\end{align}
\end{lemma}
\begin{proof}
	Note that $S:=[0,1]$ has no isolated points, and $R_a \subset S$ for every $a \subset \G$. Then recalling Remark \ref{rem:sub2main}, Proposition \ref{prop:isoInBd} implies $\Iso(R_a) \subset \Bd (R_a|S)$, and therefore $\Theta \subset \Gamma :=  \bigcup_{a \subset \G} \Bd (R_a|S)$. 
	
	Let $s \in \Theta \subset \Gamma$. Then Lemma \ref{lem:point2subset} implies there exists $i \in \{1,...,M\}$ such that \begin{equation} s \in \Bd \left( r^{-1}  \llbracket \g_i \rrbracket  \mid S \right) = \Bd \left( (g_i \circ r)^{-1} (-\infty,0] \mid S \right).\label{eq:inBoundary} \end{equation} 
	This in turn implies $g_i \circ r (s) = 0$ by Lemma \ref{lem:zeroOnBoundary}, and therefore \begin{equation}
		s \in  (g_i \circ r)^{-1}(-\infty, 0] = r^{-1}  \llbracket \g_i \rrbracket . \label{eq:ingi}
	\end{equation}
	To obtain a contradiction, suppose \eqref{eq:onBoundary} -- \eqref{eq:bounce} are all false. Since \eqref{eq:inBoundary} is true and \eqref{eq:onBoundary} false, $s \in (0,1) = \Int(S)$. Furthermore, \eqref{eq:bounce} is false, and therefore $(g_i \circ r)'(s) \neq 0$. This implies $s$ is not a local minimum of $g_i \circ r$, so for any $S$-neighbourhood $V$ of $s$, there exists $\hat{s} \in V$ such that $g_i \circ r(\hat{s}) \leq g_i \circ r(s) = 0$. Therefore \begin{equation} \label{eq:notIsoI} s \notin \Iso \big( (g_i \circ r)^{-1}(-\infty, 0] \big) =  \Iso \left( r^{-1} \llbracket \g_i \rrbracket  \right). \end{equation} Since $s \in \Theta$, there exists $a \subset \G$ such that \begin{equation} s \in \Iso (R_a) \subset R_a. \label{eq:inrHa} \end{equation} Now recall \eqref{eq:preimageHa}, which is restated below: 
	\begin{equation*} R_a = \left( \bigcap_{\g \in a} r^{-1} \llbracket \g \rrbracket  \right) \cap \left( \bigcap_{\g \in \G \setminus a} \left( r^{-1} \llbracket \g \rrbracket \right)^c\right).  \end{equation*}
	From \eqref{eq:ingi} it follows that $\g_i \in a$, because if $\g_i \notin a$, then $s \in (r^{-1} \llbracket \g_i \rrbracket)^c$. The above equation can therefore be re-written as
	\begin{equation} R_a = r^{-1} \llbracket \g_i \rrbracket  \cap G_a, \label{eq:rInt} \end{equation}
	where
	$$ G_a:= \left( \bigcap_{\g \in a\setminus\{\g_i\}} r^{-1} \llbracket \g \rrbracket  \right) \cap \left( \bigcap_{ \g \in \G \setminus a} \big( r^{-1}  \llbracket \g \rrbracket \big)^c\right).$$
	Given \eqref{eq:rInt}, Lemma \ref{lem:isoMinus} implies
	\begin{equation} \Iso(R_a) \setminus \Bd(G_a \mid S) \subset \Iso \left( r^{-1} \llbracket \g_i \rrbracket  \right), \label{eq:isoMinusBd} \end{equation}
	and applying Proposition \ref{prop:BdIntersection} to the definition of $G_a$, 
	\begin{align*} \Bd(G_a|S) & \subset \left( \bigcup_{\g \in a \setminus \{\g_i\}} \Bd( r^{-1} \llbracket \g \rrbracket  \mid S) \right) \cup \left( \bigcup_{\g \in \G \setminus a} \Bd \left(\big(r^{-1}  \llbracket \g \rrbracket \big)^c \mid S \right) \right)\\
		& = \bigcup_{\g \in \G \setminus \{\g_i\}} \Bd( r^{-1} \llbracket \g \rrbracket \mid S).\end{align*}
	Since \eqref{eq:doubleCrossing} is also false, $s \notin \Bd \left( r^{-1} \llbracket \g_j \rrbracket \mid S \right)$ for every $j \neq i$. Then clearly $s \notin \Bd(G_a|S)$, and together with \eqref{eq:inrHa}, \eqref{eq:isoMinusBd} implies $s \in \Iso \left( r^{-1} \llbracket \g_i \rrbracket \right)$, which contradicts \eqref{eq:notIsoI}.
\end{proof}
\begin{corollary}[Necessary conditions for isolated points] \label{cor:3cond}
	Under the hypotheses of Lemma \ref{lem:IsoPoints}, let
	$$\mathcal{C}:=  
	\left\{i \in \{1,...,M\} \mid \{0\} \subsetneq g_i \circ r \big[[0,1]\big]  \right\}.$$ If $s \in \Theta$ then at least one of the following hold:
	\begin{enumerate}[label={\roman*)}]
		\item $s \in \{0,1\}$ and there exists $i \in \mathcal{C}$ such that $g_i \circ r(s) = 0$ \label{pt:onBoundary}
		\item there exist $i,j \in \mathcal{C}$ with $i \neq j$ such that $g_i \circ r(s) = g_j \circ r(s) = 0$\label{pt:doubleCrossing}
		\item there exists $i \in \mathcal{C}$ such that $g_i \circ r(s) = (g_i \circ r)'(s) = 0$. \label{pt:bounce}
	\end{enumerate}
\end{corollary}
\begin{proof}
	Applying \eqref{eq:giRelship} and Lemma \ref{lem:zeroOnBoundary} to \eqref{eq:onBoundary} -- \eqref{eq:bounce} yields \ref{pt:onBoundary} -- \ref{pt:bounce}, respectively, but for $1 \leq i,j \leq M$. Corollary \ref{cor:zeroPoly} additionally confirms that $i,j \in \mathcal{C}$.  
\end{proof}
\begin{remark}
	Consider the case in Remark \ref{rem:isolatedPoints} for $z = h \circ r$. Equations \eqref{eq:rem0} -- \eqref{eq:rem1} imply $s^*_m \in (0,1)$, and Remark \ref{rem:isoTheta} implies $s^*_m \in \Theta$. Corollary \ref{cor:3cond} now makes it clear that, for $s = s^*_m$, \ref{pt:doubleCrossing} and \ref{pt:bounce} are the only possible conditions under which \eqref{eq:rem0} -- \eqref{eq:rem1} occur. Specifically, $s^*_m$ is a root of two distinct non-zero $g_i \circ r$, or a shared root of some non-zero $g_i \circ r$ with its derivative. Thus, potential isolated points can be identified by testing for the presence of double roots. 
\end{remark}
In view of Figure \ref{fig:isoPoints}, we refer to points satisfying \eqref{eq:doubleCrossing} and \eqref{eq:bounce} as \emph{double crossing} and \emph{bounce} points, respectively. 

\subsection{Algebraic conditions on sampling functions}
We are now ready to state the main result, which specializes Theorem \ref{thm:topological} to the setting described by Assumption \ref{ass:main}.

\begin{theorem} \label{thm:polynomials}
	Suppose Assumption \ref{ass:main} holds. Define
	\begin{equation} \mathcal{C}:=  
		\left\{ i \in \{1,...,M\} \mid \{0\} \subsetneq g_i \circ r \big[[0,1]\big]  \right\},
		\label{eq:Cn} \end{equation}
	let $P \in \R[s]$ be a non-zero polynomial such that
	\begin{equation}
		\prod_{i \in \mathcal{C}} g_i \circ \rho \mid P, \label{eq:Pn}
	\end{equation}
	and $ V:= \gcd(P,P')$. If $\sigma:\K \to [0,1]$ is a finite sampling function such that \begin{align}
		& V^{-1}\{0\} \cap [0,1] \subset \sigma[\K], \label{eq:isoFrame} \\
		\forall k \in \K^\ominus,\ \big|& P^{-1}\{0\} \cap [\sigma(k), \sigma(k+1)]\big| \leq 1, \label{eq:singleChange2}
	\end{align} 
	then $h \circ r \circ \sigma $ is a trace of $h \circ r$.
\end{theorem}
Once again, we pause for interpretation before proceeding with the proof. Recall that $r$ is the restriction of the vector polynomial $\rho$ to $[0,1]$. 
Equation \eqref{eq:Cn} can therefore be expanded to
\begin{equation} \mathcal{C} := \{ i \in \{1,...,M\} \mid g_i \circ \rho \neq 0 \ \land \ \exists s \in [0,1],\ g_i \circ \rho(s) = 0\}. \label{eq:CnLong} \end{equation}
Thus $\mathcal{C}$ selects all non-zero $g_i \circ \rho$ with roots in $[0,1]$ to be factors of $P$ in \eqref{eq:Pn}. Referring to Theorem \ref{thm:topological}, all boundary points in $\Gamma$ are roots of $P$. Similarly, any isolated points in $\Theta$ are repeated roots of $P$, and therefore roots of $V$. Conditions \eqref{eq:isoFrame} -- \eqref{eq:singleChange2} then ensure the conditions of Theorem 1 are satisfied. The next section considers how to construct a sampling function that satisfies \eqref{eq:isoFrame} -- \eqref{eq:singleChange2}. 
\begin{proof}
	Let $S := [0,1]$, $R_a:= r^{-1} h^{-1}\{a\}$ for $a \subset \G$, and define $\Gamma$ and $\Theta$ as in \eqref{eq:GammaTheta}. 
	First note that each $g_i \circ \rho \in \R[s]$, because $\rho \in \R[s]^n$ and $g_i \in \R[x_1,...,x_n]$ by Assumption \ref{ass:main}. 
	The following chain of equalities and inclusions is now established:
	\begin{align} \Gamma & \subset \bigcup_{\g \in \G} \Bd \big( r^{-1}\llbracket \g \rrbracket \mid S \big) = \bigcup_{i=1}^M \Bd \left( g_i \circ r^{-1}(-\infty,0]  \mid S \right)  = \bigcup_{i \in \mathcal{C}}  \Bd \left( g_i \circ r^{-1}(-\infty,0]  \mid S \right) \label{eq:gammaRewrite}\\ 
		& \subset \bigcup_{i \in \mathcal{C}} g_i \circ r^{-1}\{0\} \subset \bigcup_{i \in \mathcal{C}} (g_i \circ \rho)^{-1}\{0\} \subset P^{-1}\{0\} . \label{eq:PinvInclusion} 
	\end{align}
	The first inclusion of \eqref{eq:gammaRewrite} follows from Lemma \ref{lem:point2subset}, and \eqref{eq:giRelship} establishes the subsequent equality. For any $1 \leq i \leq M$, if $i \notin \mathcal{C}$, the contrapositive of Corollary \ref{cor:zeroPoly} implies $\Bd \left( g_i \circ r^{-1}(-\infty,0]  \mid S \right) = \emptyset$, which in turn establishes the second equality of \eqref{eq:gammaRewrite}. The first inclusion of \eqref{eq:PinvInclusion} follows from Lemma \ref{lem:zeroOnBoundary}, and the second holds because $r$ is a restriction of $\rho$. If $g_i \circ \rho(s) = 0$ for $i \in \mathcal{C}$ and $s \in \R$, then \eqref{eq:Pn} implies $P(s) = 0$, which establishes the final inclusion.
	If \eqref{eq:singleChange2} is satisfied, then \eqref{eq:gammaRewrite} -- \eqref{eq:PinvInclusion} imply Condition \ref{cond:gamma} of Theorem \ref{thm:topological} is also satisfied. Also since $g_i \circ \rho$ is non-zero for all $i \in \mathcal{C}$, $P$ can indeed be non-zero, as assumed.
	
	Now suppose $s^\star \in \Theta \subset [0,1]$. Then at least one of \ref{pt:onBoundary} -- \ref{pt:bounce} of Corollary \ref{cor:3cond} hold true. If \ref{pt:onBoundary}, then $s^\star \in \{0,1\} \subset \sigma[\K]$ by Definition \ref{def:finite_sampling}. Consider now the latter two points. If \ref{pt:doubleCrossing} holds, then $g_i \circ \rho(s^\star) = g_j \circ \rho(s^\star)=0$ for some distinct $i,j \in \mathcal{C}$, and it follows from \eqref{eq:Pn} that $(s-s^\star)^2\mid P$. If \ref{pt:bounce}, then $g_i \circ \rho(s^\star) = (g_i \circ \rho)'(s^\star) = 0$ for some $i \in \mathcal{C}$, which implies $(s-s^\star)^2 \mid g_i \circ \rho$ by Proposition \ref{prop:repeatedRoots}, which in turn implies $(s-s^\star)^2\mid P$. In either case then, $(s-s^\star)^2\mid P$. Now $Q:=P/V$ is square-free by Proposition \ref{prop:mignotte}, and $P = QV$.  If $(s-s^\star)\nmid V$, then $(s-s^\star)^2 \mid Q$, which is a contradiction. Thus $V(s^\star) = 0$, and $s^\star \in V^{-1} \{0\} \cap [0,1]$. If \eqref{eq:isoFrame} holds, then $s^\star \in \sigma(\K)$. It has been shown that $\Theta \subset \sigma[\K]$, satisfying Condition \ref{cond:theta} of Theorem \ref{thm:topological}. 
\end{proof}
\begin{remark} \label{rem:flex}
	Condition \eqref{eq:Pn} permits some flexibility in the choice of $P$, which is useful for the algorithm designer. This issue is discussed further in Section \ref{sec:complexity}, but it is worth noting that there is no clear choice of $P$ that yields optimal computational complexity for all problem instances.  
\end{remark}
\begin{remark}
Throughout this section, it has not been necessary to explicitly assume finite variability in any of the results. In fact, Assumption \ref{ass:main} guarantees that $h \circ x$ is of finite variability, because all univariate polynomials have a finite number of roots. 
\end{remark}
\section{Algorithmic implementation} \label{sec:numerical}

\subsection{Overview of sampling strategy} \label{sec:strategyOutline}
According to Theorem \ref{thm:polynomials}, we can compute a trace of $h \circ r$ by constructing a sampling function that places checkpoints
\begin{enumerate}[label={\arabic*})]
	\item on either side of every root of $P$ in $[0,1]$ to satisfy \eqref{eq:singleChange2} \label{step:gamma}; and
	\item exactly at every root of $V$ in $[0,1]$ to satisfy \eqref{eq:isoFrame}. \label{step:theta}
\end{enumerate}
Since $P$ and $V$ are univariate polynomials, we can exploit some concepts and subroutines from Computer Algebra to do this algorithmically. In particular, we require a root existence test and a root isolation algorithm. 

Before proceeding with a discussion of these subroutines, recall that any polynomial $p \in \Q[s]$ with rational coefficients can be scaled to form a polynomial $q \in \mathbb{Z}[s]$ with integer coefficients, without changing its roots. Specifically, let $\textsc{RationalToInteger}(p) = q$ multiply $p$ by the product of the denominators of its coefficients. Its arithmetic complexity is $O(\deg(p)^2)$. 
\subsection{Root existence test} \label{sec:rootExistence}
The result below, based on Descartes' Rule of Signs, provides a computationally robust test to determine whether a univariate polynomial has any real roots in an open interval.  
\begin{proposition}
	Let $p \in \R[s]$ be non-zero, $d := \deg(p)$, and \begin{equation} q(s):= (s+1)^d p \left( \frac{as+b}{s+1} \right), \label{eq:q} \end{equation}
	where $a<b$. All the non-zero coefficients of $q$ have the same sign iff $p$ has no roots in the interval $(a,b)$.
\end{proposition}
\begin{proof}
	As noted in \cite[Section 3]{mehlhornDeterministicAlgorithmIsolating2011}, the M\"{o}bius transformation $\left(s \mapsto \frac{as+b}{s+1}\right)$ is a bijection from $(0,\infty)$ to $(a,b)$. Furthermore, $q \in \R[s]$. Thus for any $s>0$, $\big( q(s) = 0 \iff p(\hat{s}) = 0\big)$, where $\hat{s} := \frac{as+b}{s+1} \in (a,b)$. Let $\var(q)$ denote the number of sign changes in the sequence of coefficients of $q$~\cite{mehlhornDeterministicAlgorithmIsolating2011}.
	If $\var(q) = 0$, then $q$ has no roots in $(0,\infty)$ by \cite[Theorem 4]{collinsRealZerosPolynomials1982}, which implies $p$ has no roots in $(a,b)$. Conversely, if $p$ has no roots in $(a,b)$, then $q$ has no roots in $(0,\infty)$, which implies $\var(q) = 0$ by \cite[Theorem 5]{collinsRealZerosPolynomials1982}.
\end{proof}  

Suppose $p \in \R[s]$ is given by $p(s) = \sum_{i=0}^d p_i s^i \in \R[s]$, and consider the following operators of type $\R[s] \to \R[s]$ for $s,\lambda \in \R$:  
\begin{align*}
	(\mathsf{R}p)(s) & := \sum_{i=0}^d p_{d-i}s^i = \begin{cases} s^d p\left(s^{-1} \right),& \IF s \neq 0 \\
		p_d,& \IF s = 0 \end{cases},  \\
	(\mathsf{T}_\lambda p)(s) & := \sum_{i=0}^d \left( \sum_{j=i}^d  {j \choose i} p_j \lambda^{j - i} s^i \right) = p(s + \lambda), \\
	(\mathsf{C}_\lambda p)(s) & := \sum_{i=0}^d p_i \lambda^i s^i = p(\lambda s).
\end{align*}  The transformed polynomial $q$ in \eqref{eq:q} can then be computed according to
$ q  = \mathsf{T}_1 \mathsf{RC}_{b-a} \mathsf{T}_a p.$
Operator $\mathsf{R}$ is the \emph{reciprocal transformation}~\cite[Definition 4, Remark 6]{krandickNewBoundsDescartes2006}, which simply reverses the order of the coefficients.
The \emph{Taylor shift}~\cite[Chapter 4.1]{gerhardModularAlgorithmsSymbolic2005} operation $\mathsf{T}_\lambda$ can be performed in $O(d^2)$ arithmetic operations \cite[Theorem 4.3]{gerhardModularAlgorithmsSymbolic2005}. 

\begin{remark} The existence of roots of $p \in \Q[s]$ in the interval $(a,b)$ can be determined in $O( \deg(p)^2)$ arithmetic operations. This can obviously be extended to the closed interval by evaluating $p(a)$ and $p(b)$. \label{rem:rootExistence} \end{remark}
\subsection{Root isolation subroutine} \label{sec:rootIsolation}
Given a polynomial $p \in \R[s]$ with root $s^\star \in \R$, a bounded interval is an \emph{isolating interval} for $s^\star$ iff it is either open or a singleton, contains $s^\star$, and contains no other roots of $p$. An isolating interval for $s^*$ is a \emph{strict isolating interval} iff it is open and its closure contains no other roots of $p$. 

We focus below on constructing isolating intervals for the roots of polynomials in $(0,1)$. For computational reasons, attention is restricted to polynomials with rational coefficients and intervals with rational endpoints.  
\begin{definition}[Root isolation algorithm] \label{def:weakRootIsolation}
	Given $p \in \Q[s]$, a root isolation algorithm returns a set of intervals \begin{equation} \mathcal{J} := \{ I_j \subset (0,1)  \mid 1 \leq i \leq J\}, \label{eq:typicalIsolatingInterval} \end{equation} 
	of cardinality $J \leq \deg(p)$, 
	such that all of the following hold:
	\begin{enumerate}[label={\theenumi)}]
		\item each $I_j$ is either open or a singleton
		\item each $I_j$ has rational endpoints
		\item $\sup I_j \leq \inf I_{j+1}$ for all $1 \leq j < J$
		\item every root of $p$ in $(0,1)$ is contained in some $I_j$
		\item every $I_j$ contains exactly one root of $p$. 
	\end{enumerate}
\end{definition}
The output \eqref{eq:typicalIsolatingInterval} is represented in pseudocode by $I_{1:J} = $\textsc{RootIsolation}$(p)$. 
Clearly each $I_j \in \mathcal{J}$ is an isolating interval for some root of $p$. Letting $a_j := \inf I_j$ and $b_j := \sup I_j$ for $1 \leq j \leq J$, the interval endpoints satisfy
$$0 \leq a_1 \leq b_1 \leq a_2 \leq b_2 \leq \hdots \leq a_J \leq b_J \leq 1.$$ The Vincent, Collins and Akritas (VCA) Algorithm~\cite{collinsPolynomialRealRoot1976}, also known as the Modified Uspensky or Descartes method, is a well-studied root isolation algorithm that performs a bisection search using Descartes Rule of Signs. Its output complies with Definition \ref{def:weakRootIsolation}, and the algorithm is presented concisely in \cite[Algorithm 1]{mehlhornDeterministicAlgorithmIsolating2011}. 
\begin{remark} \label{rem:rootIsolation} If $p \in \mathbb{Z}[s]$ is square-free, then the bit complexity of the VCA algorithm is $O(d^5(L + \log d)^2)$, where $d = \deg(p)$ and $\|p\|_\infty < 2^L$~\cite[Theorem 4.1]{eigenwilligAlmostTightRecursion2006}. \end{remark}
Our strategy is to use a root isolation algorithm to place a checkpoint on either side of the roots of $P$. However, Definition \ref{def:weakRootIsolation} permits roots at the boundaries of the isolating intervals, which leads to problems. In order to guarantee \eqref{eq:singleChange2}, all the isolating intervals must be strict. We therefore adopt the more stringent definition below. Note that singleton intervals are not permitted. 
\begin{definition}[Strict root isolation algorithm] \label{def:rootIsolation}
	Given any $p \in \Q[s]$, a strict root isolation algorithm returns a set of intervals $$ \mathcal{J}^* := \{ (a_j,b_j) \subset (0,1) \mid 1 \leq i \leq J\},$$ of cardinality $J \leq \deg(p)$, such that all of the following hold:
	\begin{enumerate}[label={\theenumi)}]
		\item $a_j, b_j \in \Q$ for all $1 \leq j \leq J$
		\item $0 \leq a_1 < b_1 \leq a_2 < b_2 \leq \hdots \leq a_J < b_J \leq 1$  \label{cl:isoIntervalEndpointOrdering}
		\item every root of $p$ in $(0,1)$ is contained in some interval $(a_j,b_j)$
		\item every interval $(a_j,b_j)$ contains exactly one root of $p$
		\item $p(a_1) \neq 0 \neq p(b_J)$. \label{cl:isoIntervalEndpointsOnBoundary}
	\end{enumerate}
\end{definition}
\begin{remark} \label{rem:endpointRoots}
	It follows from Definition \ref{def:rootIsolation} that. for all $1 \leq j \leq J$, the closed interval $[a_j,b_j]$ contains exactly one root of $p$. Also observe there is at most one root of $p$ in $[0,a_1]$, which can only occur if $p(0) = 0$. Similarly, there is at most one root of $p$ in $[b_J,1]$, which only occurs when $p(1) = 0$. The proof of Theorem \ref{thm:alg} in Section \ref{sec:splineTrace} relies on these properties to establish \eqref{eq:singleChange2}. 
\end{remark}

Using lower bounds on polynomial root separation \cite{collinsPolynomialMinimumRoot2001}, the output of a root isolation algorithm is readily modified into the output of a strict root isolation algorithm. 
Algorithm \ref{alg:stricten}, \textsc{StrictIsolatingIntervals}, offers a procedure for doing this. Since this procedure is somewhat distracting, and motivated by aspects of the main algorithm yet to be presented, the details are deferred to Algorithm \ref{alg:stricten} in Appendix \ref{app:stricten}.
\subsection{\textsc{PolyTrace} algorithm} \label{sec:splineTrace}

Recall the two-step procedure outlined in Section \ref{sec:strategyOutline}. 
Step \ref{step:gamma} can be achieved by means of a root isolation algorithm.
Step \ref{step:theta} is problematic, because in general the roots of $V$ cannot be computed exactly. And even if they could, the roots may not be rational. Approximate roots will not do, because the roots of $V$ extract isolated points, which by definition capture information absent from their neighbouring points. 
The following result demonstrates that, even if an isolated point $s^\star \in [0,1]$ is not known precisely, the set $h \circ r(s^\star)$ needed to construct a trace can still be evaluated by means of a root existence test. Only a strict isolating interval for $s^\star$ is required. 
\begin{lemma}[Isolation Lemma] \label{lem:sampleIsolated}
	Under the Assumptions of Theorem \ref{thm:polynomials}, suppose there exist $0 \leq a < s^\star < b \leq 1$ such that
	\begin{equation} \{ s \in [a,b] \mid P(s) = 0 \} = \{ s^\star\}. \label{eq:singleRootQ} \end{equation}
	Then $h \circ r(s^\star) = h \circ r(a)  \cup \{ \g_i \in \G \mid \exists \hat{s} \in (a,b),\ g_i \circ \rho(\hat{s}) = 0 \}.$
\end{lemma}
The Isolation Lemma tells us that, given a strict isolating interval $(a,b)$ for the root $s^\star \in [0,1]$ of $P$, the set $h \circ r(s^\star)$ can be computed by evaluating $h\circ r(a)$, and then adding to it all $\g_i$ for which $g_i \circ \rho$ has a root in $(a,b)$. 
\begin{proof}
	Define $E:=\{ \g_i \in \G \mid \exists \hat{s} \in (a,b),\ g_i \circ \rho(\hat{s}) = 0 \}$. It is first shown that (i) $ E\subset h \circ r(s^\star)$, then that (ii) $h \circ r(a) \cup E \subset h \circ r(s^\star)$, and finally that (iii) $h \circ r(s^\star) \subset h \circ r(a) \cup E$. 
	
	Observe that 
	\begin{equation}
		\forall s \in [0,1],\	h \circ r(s) = h \circ \rho(s) = \{ \g_i \in \G \mid g_i \circ \rho(s) \leq 0\}. \label{eq:hstar}
	\end{equation}
	(i) Suppose $\g_i \in E$. Then $g_i \circ \rho(\hat{s}) = 0 $ for some $\hat{s} \in (a,b)$. 
	Recall \eqref{eq:CnLong}, restated below:
	\begin{equation} \mathcal{C} := \{ i \in \{1,...,M\} \mid g_i \circ \rho \neq 0 \ \land \ \exists s \in [0,1],\ g_i \circ \rho(s) = 0\}. \tag{\ref{eq:CnLong}} \end{equation}
	If $i \notin \mathcal{C}$, then \eqref{eq:CnLong} implies $g_i \circ \rho = 0$ and therefore $g_i \circ \rho(s^\star) = 0$, which in turn implies $\g_i \in h \circ r(s^\star)$ by \eqref{eq:hstar}. Suppose instead that $i \in \mathcal{C}$. Then $g_i \circ \rho \mid P$ by \eqref{eq:Pn}, and furthermore $P(\hat{s}) = 0$. The assumption \eqref{eq:singleRootQ} then implies $\hat{s} = s^\star$, by which $g_i \circ \rho(s^\star) = 0$ and therefore $\g_i \in h \circ r(s^\star)$. Thus $ E \subset h \circ r(s^\star)$.
	
	(ii) Suppose now that $\g_i \in h \circ r(a)$, which implies \begin{equation} g_i \circ \rho(a) \leq 0. \label{eq:aInH} \end{equation}
	If $i \notin \mathcal{C}$, then by \eqref{eq:CnLong} either $g_i \circ \rho = 0$, which implies $g_i \circ \rho(s^\star) = 0$, or $g_i \circ \rho$ has no roots in $[0,1]$, which implies $g_i \circ \rho(a) < 0$, and moreover, that $g_i \circ \rho(s^\star)<0$ due to continuity. In either case, $g_i \circ \rho(s^\star) \leq0$, hence $\g_i \in h \circ r(s^\star)$. Suppose instead that $i \in \mathcal{C}$. Then $g_i \circ \rho \mid P$ by \eqref{eq:Pn}, and since $s^\star \in (a,b)$, \eqref{eq:singleRootQ} implies $P(a) \neq 0$. Since $g_i \circ \rho \mid P$, it follows that $g_i \circ \rho(a) \neq 0$, and therefore $ g_i \circ \rho(a) < 0$ by \eqref{eq:aInH}. If $g_i \circ \rho(s^\star) \leq 0$, then $\g_i \in h \circ r(s^\star)$ by \eqref{eq:hstar}. If $g_i \circ \rho(s^\star) > 0$, then by the Intermediate Value Theorem there exists $\hat{s} \in (a,s^\star) \subset (a,b)$ such that $g_i \circ \rho(\hat{s}) = 0$, and therefore $\g_i \in E \subset h \circ r(s^\star)$. Thus $h \circ r(a) \subset h \circ r(s^\star)$, which establishes that $h \circ r(a) \cup E \subset h \circ r(s^\star)$. 
	
	(iii) Finally, suppose $\g_i \in h \circ r(s^\star)$, which implies $g_i \circ \rho(s^\star) \leq 0$. If $g_i \circ \rho(s^\star) = 0$, then $\g_i \in E$. Suppose now $g_i \circ \rho(s^\star) < 0$. If $g_i \circ \rho(a) \leq 0$, then $\g_i \in h \circ r(a)$. Alternatively if $g_i \circ \rho(a) > 0$, then by the Intermediate Value Theorem there exists $\hat{s} \in (a,s^\star) \subset (a,b)$ such that $g_i \circ \rho(\hat{s}) = 0$, and therefore $\g_i \in E$. Thus $h \circ r(s^\star) \subset h \circ r(a) \cup E$. 
\end{proof} 
Lemma \ref{lem:sampleIsolated} is exploited in Lines \ref{ln:IfIso1} -- \ref{ln:IfIso2} of Algorithm \ref{alg:splineTrace}, the correctness of which is now established. 
\begin{algorithm}[H] 
	\caption{Computes a trace of $h \circ \rho{\upharpoonright_{[0,1]}}$, where $\rho \in \Q[s]^n,\ g_1 \hdots g_M \in \Q[x_1,...,x_n]$,\\ and $h(x):= \{i \in \{1,...,M\} \mid g_i(x) \leq 0 \}$.} \label{alg:splineTrace}
	\begin{algorithmic}[1]	
		\Function{PolyTrace}{$\rho,\ g_{1:M}$} 
		\State $\zeta(0) := h \circ \rho(0)$ \label{ln:zeta1}
		\State $P:=1$ \label{ln:P0}
		\For{$1 \leq i \leq M$} \label{ln:formP0}
		\State $p_i := \Call{RationalToInteger}{g_i \circ \rho}$ \label{ln:pi}
		\If{$p_i$ has roots in [0,1] \textbf{and} $\deg(p_i)>0$} \label{ln:rootTest}
		\State $P \leftarrow P \times p_i$
		\EndIf
		\EndFor \label{ln:P1}
		\State $V := \Call{GCD}{P, P'}$ \label{ln:vn} 
		\State $Q := P/V$ \label{ln:qn}
		\State $I_{1:J} :=$ \Call{RootIsolation}{$Q$} \label{ln:rootIsolation}
		\For{$ 1 \leq j \leq J$}  \label{ln:Ej0}
		\State $E_j := \{ i \in \{1,...,M\} \mid p_i \text{ has roots in } I_j\}$ \label{ln:Ej}
		\EndFor \label{ln:Ej1}
		\State $(a_{1:J},b_{1:J}) := $ \Call{StrictIsolatingIntervals}{$p_{1:M}, I_{1:J}, E_{1:J}$} \label{ln:rootIsolationStrict}
		\State $k:=1$
		\For{$1 \leq j \leq J$} \label{ln:intervals1}
		\State 		$ \zeta(k) := h \circ \rho(a_j)$ \label{ln:zeta_aj}
		\State $k \leftarrow k+1$
		\If{$\deg(V) > 0$} \label{ln:IfIso1}
		\State $\zeta(k) := \zeta(k-1) \cup E_j $ \label{ln:zetaIsol} 
		\State $k \leftarrow k+1$
		\EndIf \label{ln:IfIso2}
		\EndFor \label{ln:intervals2}
		\If{$J>0$}
		\State $ \zeta(k) := h \circ \rho(b_J)$ \label{ln:zeta_bj}
		\Else
		\State $\zeta(k) := h \circ \rho\left(\frac{1}{2}\right)$ \label{ln:zetaDk}
		\EndIf
		\State $k \leftarrow k+1$
		\State $\zeta(k) := h \circ \rho(1)$ \label{ln:zetaEnd}
		\State \Return $\zeta$
		\EndFunction
	\end{algorithmic}
\end{algorithm}
\begin{theorem} \label{thm:alg}
	Suppose Assumption \ref{ass:main} is satisfied. In addition, let $\G = \{1,...,M\}$, $\rho \in \Q[s]^n$, and $g_i \in \Q[x_1,...,x_n]$ for all $1 \leq i \leq M$. Then the output $\zeta = \text{\textsc{PolyTrace}}(\rho,g_{1:M})$ of Algorithm \ref{alg:splineTrace} is a trace of $h \circ r$.  
\end{theorem}
\begin{proof}
	First note that a checkpoint is placed at $0$ and $1$ by virtue of Lines \ref{ln:zeta1} and \ref{ln:zetaEnd}, to satisfy Definition \ref{def:finite_sampling}. 
	
	Lines \ref{ln:P0} -- \ref{ln:P1} form
	\begin{equation}
		P := \prod_{i \in \mathcal{C}} g_i \circ \rho, \label{eq:PnAlg}
	\end{equation}
	where $\mathcal{C}$ is given by \eqref{eq:CnLong}. This definition clearly satisfies \eqref{eq:Pn}, and $V$ is then computed in Line \ref{ln:vn} as prescribed by Theorem \ref{thm:polynomials}. The square-free part $Q$ of $P$ is computed in Line \ref{ln:qn}. 
	Strict isolating intervals for the roots of $Q$ in $(0,1)$ are computed in Lines \ref{ln:rootIsolation} -- \ref{ln:rootIsolationStrict}. By Corollary \ref{cor:commonRoots}, these are also strict isolating intervals for the roots of $P$ in $(0,1)$, because $P$ and $Q$ share exactly the same roots.  Checkpoints are placed at the start $a_j$ of each isolating interval in Line \ref{ln:zeta_aj}, and at the end of the final isolating interval $b_J$ in Line \ref{ln:zeta_bj}. Any remaining roots of $P$ in $[0,1]$ are included in the checkpoints at $0$ and $1$.
	As noted in Remark \ref{rem:endpointRoots}, there is exactly one root of $P$ in $[a_j,a_{j+1}]$ for any $1 \leq j < J$. Furthermore, there is at most one root in $[0,a_1]$ and in $[b_J,1]$. Thus, \eqref{eq:singleChange2} is satisfied. 
	
	Line \ref{ln:IfIso1} tests for potential isolated points by checking the degree of $V$. If $V = 1$, then $V$ has no roots in $[0,1]$. Otherwise, $\deg(V)>1$, and Lines \ref{ln:IfIso1} -- \ref{ln:IfIso2} exploit Lemma \ref{lem:sampleIsolated} to evaluate $h \circ r$ at the unique root of $P$ within every strict isolating interval. Since $P=QV$, this includes all roots of $V$ in $(0,1)$. Any additional roots of $V$ in $[0,1]$ are included in the checkpoints at 0 and 1. Thus, \eqref{eq:isoFrame} is satisfied. 
	
	Finally, if no $g_i \circ \rho$ has roots in $(0,1)$, the empty set is returned in Line \ref{ln:rootIsolation}, implying that $J=0$. In this case, $\sigma(0) = 0$ and $\sigma(2) = 1$. It is still possible that $P$ has roots at both endpoints $\{0,1 \}$. Therefore, a checkpoint $\sigma(1)$ is placed at their midpoint in Line \ref{ln:zetaDk}, to ensure there is no more than one root in $[\sigma(k), \sigma(k+1)]$ for each $k \in \{0,1\}$, as required by \eqref{eq:singleChange2}. 
\end{proof}

\subsection{Computational complexity of \textsc{PolyTrace}} \label{sec:complexity}
The complexity of Algorithm \ref{alg:splineTrace} depends heavily on the subroutines chosen for the individual steps. In this section, it is assumed that for every $1 \leq i \leq M$, $g_i$ has no more than $N$ non-zero coefficients and $\deg(p_i) \leq D$, where each $p_i \in \mathbb{Z}[s]$ is formed in Line \ref{ln:pi} of Algorithm \ref{alg:splineTrace}. To understand how $N$ and $D$ can be derived from upper bounds on $\deg(\rho)$, $\deg(g_i)$, and the state space dimension $n$, refer to Appendix \ref{app:polyComp}. 
\begin{itemize}
	\item \emph{Formation of $P$:} forming each $g_i \circ \rho$ requires $O(ND^2)$ arithmetic operations (see Appendix \ref{app:polyComp}). Each root existence test in Line \ref{ln:rootTest}  requires $O(D^2)$ arithmetic operations. These are both done $M$ times. Finally, $\deg(P) \leq MD$, and therefore forming $P$ requires $O(M^2D^2)$ arithmetic operations~\cite[Lemma 3.15]{gerhardModularAlgorithmsSymbolic2005}. Overall, Lines \ref{ln:P0} -- \ref{ln:P1} require $O(MD^2(M+N))$ arithmetic operations. 
	\item \emph{G.c.d computation:} \cite[Algorithm 10.1]{basuAlgorithmsRealAlgebraic2006}, which has $O(M^2D^2)$ arithmetic complexity, performs Lines \ref{ln:vn}--\ref{ln:qn} and returns both $Q$ and $V$ with integer coefficients. Implementations with polynomial bit complexity exist \cite[Remark 10.19]{basuAlgorithmsRealAlgebraic2006}.
	See \cite[Chapters 3, 6, 11 ]{vonzurgathenModernComputerAlgebra2013} for a detailed discussion of the available algorithms for g.c.d computation, and their complexities. 
	\item \emph{Root isolation:} this step has bit complexity $O(M^5D^5(L + \log MD)^2)$ for $Q \in \mathbb{Z}[s]$ by Remark \ref{rem:rootIsolation}, where $L \in \N$ is any constant such that $\|Q\|_\infty \leq 2^L$.
	\item \emph{Strict isolating intervals:} the arithmetic complexity of Algorithm \ref{alg:stricten} is $O(MD)$. 
	\item \emph{Computing $E_1,...,E_J$}: this requires at most $MD$ root existence tests for polynomials of degree $D$ or less, yielding an overall arithmetic complexity of $O(MD^3)$ for Lines \ref{ln:Ej0} -- \ref{ln:Ej1}. 
\end{itemize}
While a more detailed analysis of the growth in polynomial coefficients sizes is required to obtain a precise estimate, the bit complexity of \textsc{PolyTrace} is clearly polynomial. The root isolation step appears likely to dominate. 

Now as noted in Remark \ref{rem:flex}, the choice $P$ affects the complexity of the algorithm. Given the above complexity estimate of the root isolation step, omitting unnecessary factors from $P$ in \eqref{eq:Pn} potentially offers significant improvements. 
The definition \eqref{eq:PnAlg} adopted by Algorithm \ref{alg:splineTrace} constructs the lowest degree polynomial that satisfies \eqref{eq:Pn}. 
However, this requires the root existence test in Line \ref{ln:rootTest}, which adds $O(MD^2)$ complexity. An alternative definition
$$ P := \prod_{ \deg( g_i \circ \rho)>0} g_i \circ \rho $$
dispenses with the root existence test, but makes no attempt to reduce $\deg(P)$. 
Both approaches have the same worst-case asymptotic complexity, because it may well be that all $g_i \circ \rho$ have roots in $[0,1]$, making the test redundant. 
If it is known \emph{a priori} that only a few boundaries are likely to be crossed within a given segment, we expect that \eqref{eq:PnAlg} leads to a net reduction in complexity.
\section{Numerical examples}
\label{sec:sim}
\subsection{Randomly generated examples}
\begin{figure}[H]
	\centering
	\includegraphics[width=\linewidth]{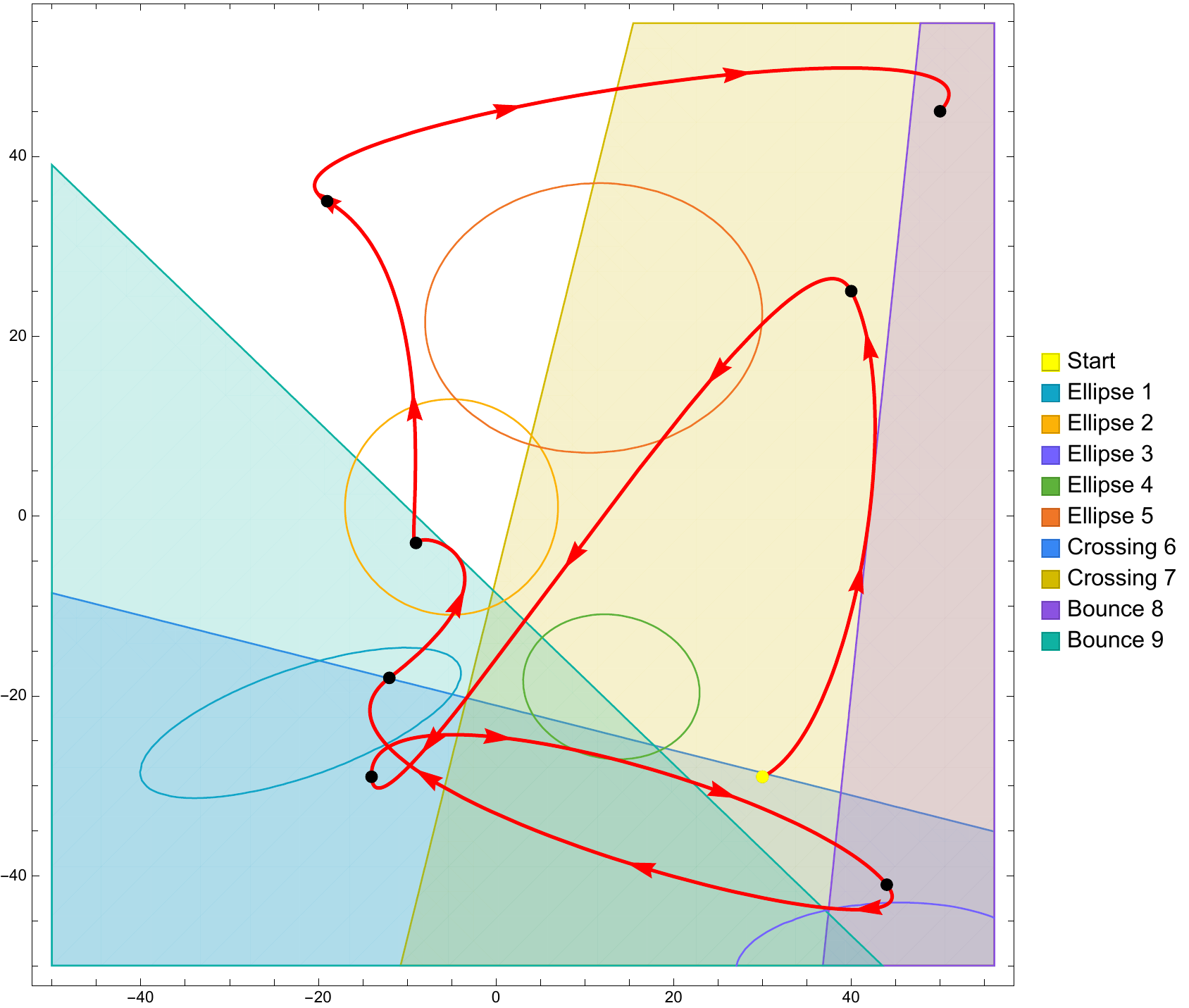}
	\begin{flushleft} \small \texttt{trace=\protect\input{testCase.txt}.}\end{flushleft}
	\caption{Polynomial spline path connecting 8 waypoints under 9 atomic propositions.}
	\label{fig:sim}
\end{figure}

Figure \ref{fig:sim} depicts a randomly generated instance of Problem \ref{prob:practical} involving a two-dimensional polynomial spline path, and atomic propositions that are mapped to either ellipses or half-spaces. Algorithm \ref{alg:splineTrace} has been implemented in Wolfram Mathematica 12.1, taking advantage of the native \textsc{RootIntervals} root isolation function. 
The output of \textsc{PolyTrace} for each segment is concatenated and included in Figure \ref{fig:sim}, after a post-processing step that removes consecutive repetitions from the trace. Region 8 has been constructed to induce a bounce point between the first and second waypoints, and Regions 6 and 7 to induce a double crossing point between the second and third waypoints. Equipped with this knowledge, the reader is invited to visually confirm the trace produced. 

\begin{figure}[H]
	\centering
	\includegraphics[width=\linewidth]{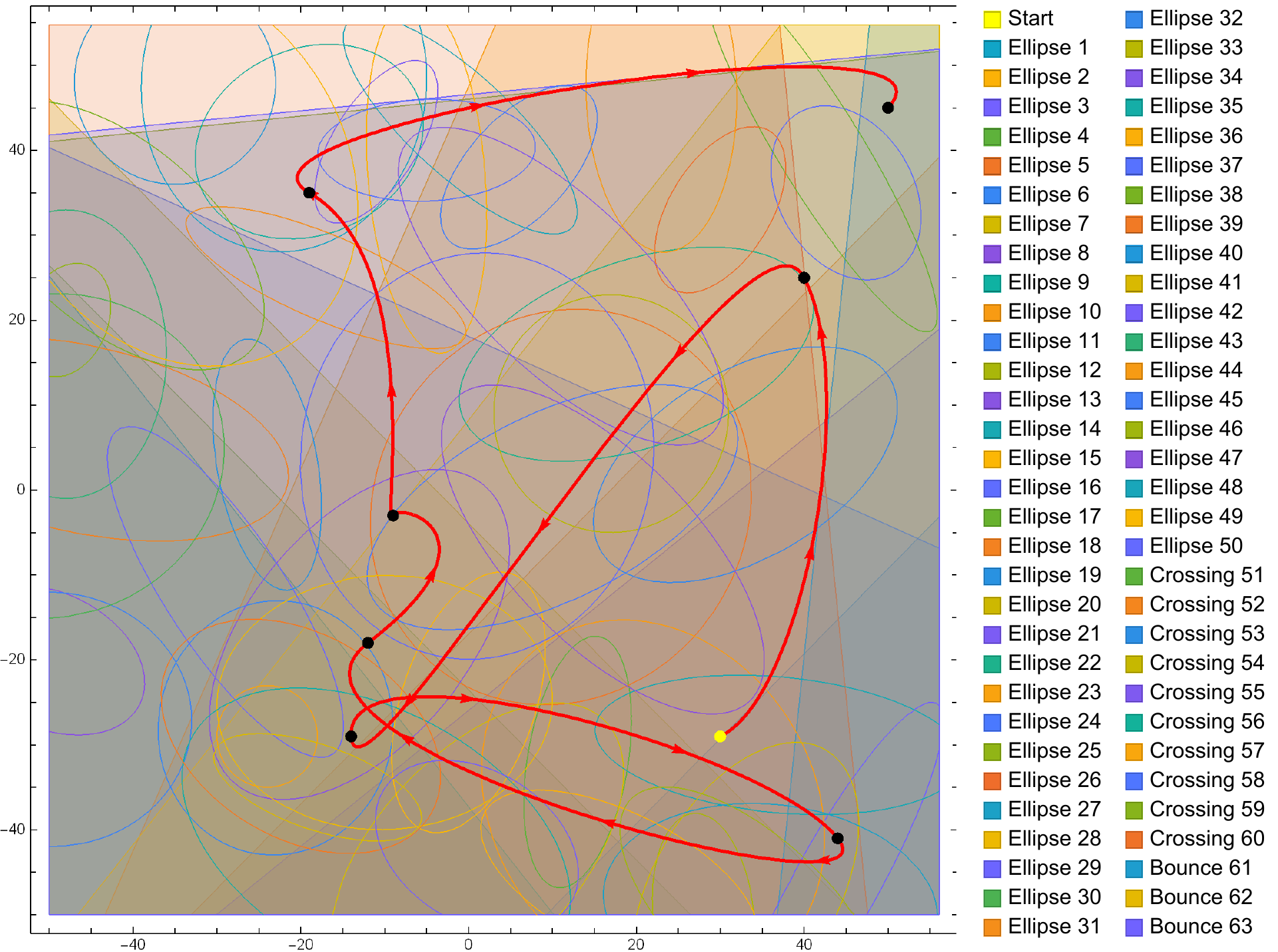}
	\begin{flushleft} \tiny \texttt{trace=\protect\input{testCase2.txt}.}\end{flushleft}
	\caption{Polynomial spline path connecting 8 waypoints under 63 atomic propositions.}
	\label{fig:sim2}
\end{figure}

The total computation time for the trace in Figure \ref{fig:sim} is 93.75 ms. Compare this with a computation time of 1.983 s for the scenario in Figure \ref{fig:sim2}, which retains the same path while increasing the number of atomic propositions sevenfold. 

\subsection{Case study: verification of a robot motion plan} \label{sec:caseStudy}
Consider a mobile robot on a factory floor, tasked with fetching a hazardous chemical compound and returning to a target zone, without colliding with obstacles. The chemical is stored in a 0.5m diameter drum, and the robot also has a circular perimeter, 1.5m in diameter. The drum attaches to the side of the robot, via a latching system. Both robot and drum must remain within a predefined 20m x 20m safe zone, which is guaranteed to be clear of moving obstacles throughout the operation. The drum, and any static obstacles, are detected using a system of cameras, and are represented within the robot's map by enclosing ellipses. The surface of the robot must make contact with the surface of the drum in order to trigger the latching system. The unstable nature of the chemical compound requires contact to be made gently, and this imposes precise requirements on the robot trajectory. Here, we encode these requirements in $\LTL_\varobslash$, and use \textsc{PolyTrace} to certify a cubic spline proposed by a path planner for the robot. The subsequent path tracking stage (see Remark \ref{rem:motionPlanning}) is guaranteed to produce only direct motions. Since this is a safety critical mission, the resulting motion plan must be verified independently, prior to execution. Here, as explained in Remark \ref{rem:motionPlanningImplications}, the verification of the motion plan depends only on the path proposed by the path planner. 

\begin{figure}[H]
	\centering
	\includegraphics[width=\linewidth]{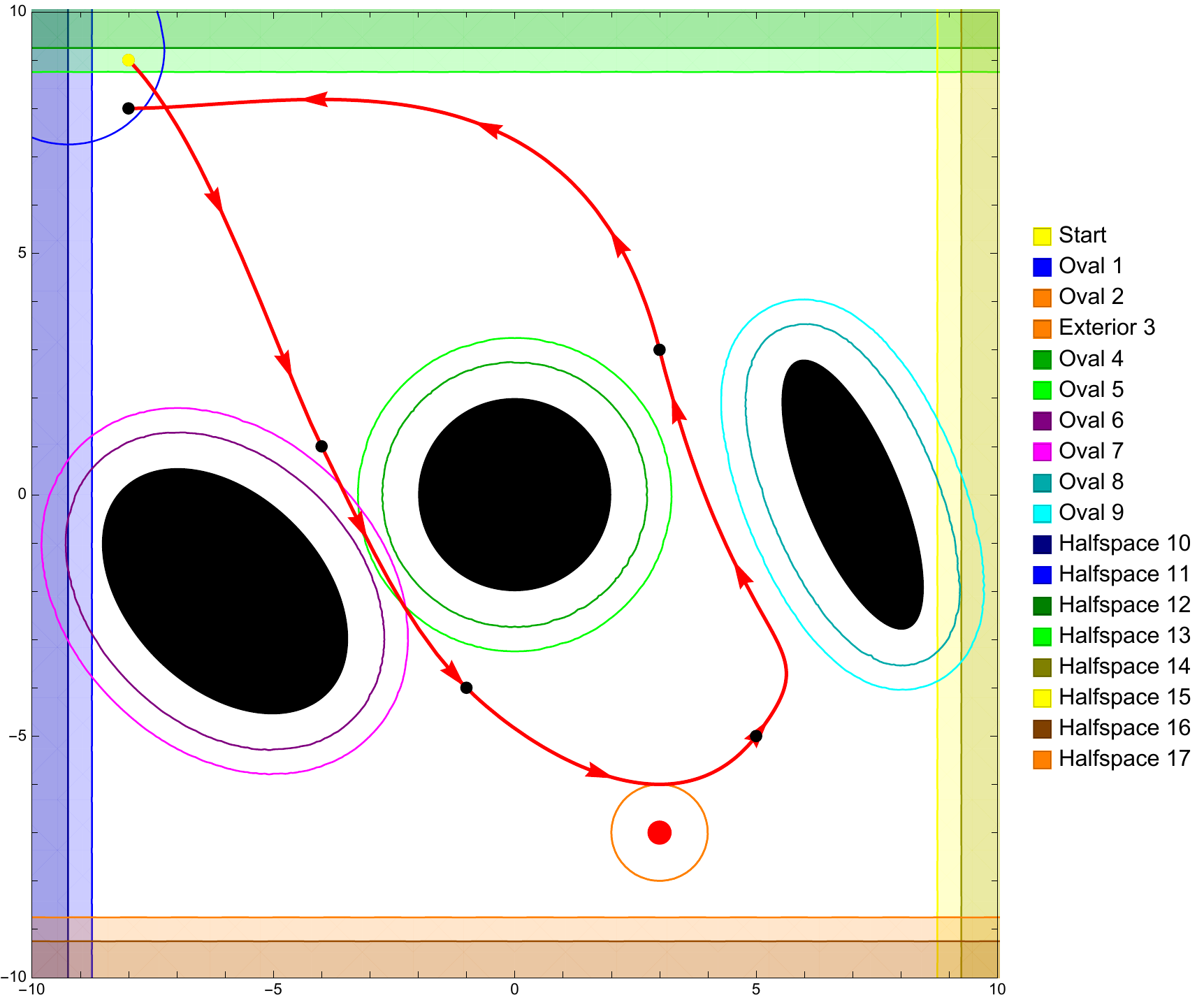}
	\begin{flushleft} \texttt{trace=\protect\input{caseStudy.txt}.}\end{flushleft}
	\caption{The proposed path for the robot is shown in red. Black filled regions represent obstacles, and the red disk at (3,-7) represents the drum. The boundaries of Oval 2 and Exterior 3 coincide. Units are in meters.}
	\label{fig:collectionBot}
\end{figure}

Figure \ref{fig:collectionBot} depicts the path superimposed on the robot's map. Obstacles are shown in solid black, the drum in red, and the safe zone corresponds exactly to the plot axes limits. We model the robot as a disk. Since the path is to be tracked by the centre of the robot, collisions can be avoided by expanding the obstacles in the map by the radius of the robot. These expanded obstacles are represented by Ovals 4, 6 and 8. In general, the expanded obstacles are no longer ellipses, but are instead defined by the 8th-order polynomial in \cite[Example 4]{sendraBriefAtlasOffset2010}. After the robot has latched onto the drum, collision avoidance for their composite shape can be achieved by further expanding the obstacles by the diameter of the drum, yielding Ovals 5, 7 and 9. Similarly, Halfspaces 10, 12, 14 and 16 enclose the safe zone contracted by the robot radius, and Halfspaces 11, 13, 15 and 17 contract it again by the drum diameter. Oval 2 expands the drum by the robot radius, so that the robot touches the drum iff its center touches Oval 2. Finally, the target zone is represented by Oval 1.

We now work with the set of atomic propositions $\G := \{\g_1,...,\g_{17} \}$. The observation map $h$ associates each $\g_i$ with the $i$th region (either Oval $i$ or Halfspace $i$) in Figure \ref{fig:collectionBot}. In order to specify the surface contact requirement, we have also defined the region `Exterior 3' by the polynomial $g_3 := - g_2 \in \Q[x_1,x_2]$. Thus, Exterior 3 is the complement of the interior of Oval 2, and the two regions only intersect on their common boundary. The formula $ \varphi_C:= \always(\g_2 \limplies \g_3)$ then permits the robot to make contact with the surface of the drum, without penetrating its interior. 
The complete mission requirement can be expressed in $\LTL_\varobslash$ as $ \varphi:= \varphi_C  \land \varphi_G \land \always \varphi_I \land \varphi_D$, where
\begin{align*} 
\varphi_G &:= \eventually(\g_2 \land \eventually \always \g_1), \\
\varphi_I &:= \lnot (\g_4 \lor \g_6 \lor \g_8 \lor \g_{10} \lor \g_{12} \lor \g_{14} \lor \g_{16}), \\
\varphi_O &:= \lnot (\g_5 \lor \g_7 \lor \g_9 \lor \g_{11} \lor \g_{13} \lor \g_{15} \lor \g_{17}), \\
\varphi_D &:= \always(\g_2 \limplies \always \varphi_O).
 \end{align*}
The mission goal is expressed by $\varphi_G$, which requires the robot to eventually make contact with the drum, before entering the target zone and remaining there. Formula $\varphi_I$ prohibits the robot from colliding with the inner boundaries of the obstacles and `unsafe' zone, which have been expanded by the robot radius. Formula $\varphi_O$ does the same for the outer boundaries, which have been expanded by the drum diameter as well. Formula $\varphi_D$ only requires $\varphi_O$ to hold from the point of contact with the drum. 

The trace of the proposed path, computed by \textsc{PolyTrace}, is given at the bottom of Figure \ref{fig:collectionBot}. This path $r:[0,1] \to \R^2$ can then be composed with a direct motion $u:[0,\infty) \to [0,1]$ to generate the direct trajectory $x = r \circ u$. 
Invoking Proposition \ref{prop:direct}, the resulting trace is
$$ h \circ x \circ \sigma = \{\g_1, \g_3, \g_{13}\}, \{\g_1, \g_3\}, \{\g_3\}, \{\g_3, \g_7\}, \{\g_3, \g_5, \g_7\}, \{\g_3, \g_7\}, \{\g_3\}, \{\g_2, \g_3\}, \{\g_3\}, \{\g_1, \g_3\}^ \omega,$$
where $\sigma$ is the corresponding sampling function. It is straightforward to verify that $h \circ x \circ \sigma \models \varphi$, which holds for any direct motion $u$ produced by the path tracking stage.

\begin{remark}
The surface contact requirement $\varphi_C$ is an example of an $\LTL_\varobslash$ formula that can only be satisfied with zero robustness. To the best of the authors' knowledge, there are no other approaches in the literature capable of verifying a continuous-time trajectory against such a specification, because the existing verification approaches rely on strictly positive robustness margins. For a real-world implementation, it is, of course, acceptable to relax $\varphi_C$ by expanding the contact boundary into a contact area. In practice, a sufficiently small expansion does not jeopardise the mission, because the use of feedback control makes the latching system robust to small deviations from the ideal path. However, the existing approaches, which are based on ~\cite{fainekosRobustnessTemporalLogic2009}, still demand high frequency sampling at a rate that is not known beforehand. The tighter the expansion, the harder it becomes to verify any path proposed. As discussed in the introduction, multiple attempts may be made without guarantee of an outcome. By contrast, even in the most extreme cases, \textsc{PolyTrace} never fails to return a conclusive result, as this example serves to demonstrate.
\end{remark}

\section{Conclusion}
\label{sec:conclusion}

This paper culminates in an algorithmic solution to a special case of Problem \ref{prob:practical}, in which the path is a polynomial, and the observation map generates semi-algebraic regions of interest. 
First, the more general problem of verifying a trajectory under an observation map against a temporal logic specification is formally stated as Problem \ref{prob:general}. Its formulation relies on the MITL semantics of \cite{alurBenefitsRelaxingPunctuality1996}, which can be interpreted over continuous-time signals of finite variability via their TSS representations. The trace of such a signal is identified with the first component of its TSS representation. The specification language $\LTL_\varobslash$ is then introduced as a fragment of MITL, and Remark \ref{rem:TSS2LTL} emphasizes that for $\LTL_\varobslash$ formulas, MITL satisfaction of a continuous-time signal can be inferred from LTL satisfaction of its trace. Narrowing the scope to $\LTL_\varobslash$ formulas, the focus of the paper shifts to Problem \ref{prob:main}: generating a trace of a trajectory. As argued in Remark \ref{rem:practicalTrajectories}, the only trajectories that are practically amenable to path checking have infinite traces that are lasso words composed of the finite traces of paths. Thus, we arrive at Problem \ref{prob:practical}: generating the trace of a path. This can be done by sampling the path judiciously, and Theorem \ref{thm:topological} presents topological conditions for a sampling function to achieve this. Section \ref{sec:concrete} applies this result to sample polynomial paths under observation maps that only generate semi-algebraic regions of interest, deriving more concrete conditions in terms of the roots of univariate polynomials. The \textsc{PolyTrace} algorithm proposed in Section \ref{sec:numerical} invokes root isolation and root existence subroutines from computer algebra to satisfy these conditions. One novel feature of this algorithm is its ability to sample at the isolated points characterised in Corollary \ref{cor:3cond}. Correctness of the algorithm is proved, and its complexity shown to be polynomial in the number of atomic propositions. Numerical examples and a case study are provided in Section \ref{sec:sim}, including an example of an LTL specification that cannot be verified using other existing methods.  

An obvious extension to our work is to relax the restriction to $\LTL_\varobslash$ specifications by permitting bounded temporal operators. In this case, satisfaction no longer depends exclusively on the trace, so new analysis is required. The precise locations of the polynomial roots are then of greater importance, because they correspond to crossing times. Furthermore, we expect the polynomial assumptions to simplify the calculation of robustness margins, and improve on the margin estimates in \cite{fainekosRobustnessTemporalLogic2009,donzeRobustSatisfactionTemporal2010, donzeEfficientRobustMonitoring2013, deshmukhRobustOnlineMonitoring2017}. This paper has only considered the verification problem, but it is possible that the synthesis problem (i.e., generating polynomial trajectories guaranteed to satisfy formal specifications) may also benefit from the methods we have discussed. 

\section{Acknowledgements}
This research has received funding from the Australian Government through the Defence Cooperative Research Centre for Trusted Autonomous Systems. Other project partners include BAE Systems Australia, DST Group, The University of Adelaide, and The University of Melbourne. The authors are grateful for their partnership, which has sparked many fruitful discussions. 
\bibliography{../References/FormalMethodsInControl.bib}
\bibliographystyle{ieeetr}
\begin{appendices}
	\section{Adjustment of non-strict isolating intervals} \label{app:stricten}
	Here we describe a procedure for adjusting the output of a root isolation algorithm to comply with Definition \ref{def:rootIsolation}. Let $a_j := \inf I_j$ and $b_j := \sup I_j$ for all $I_j \in \mathcal{J}$, where $\mathcal{J}$ complies with Definition \ref{def:weakRootIsolation}. There are two problems to be dealt with, both involving roots at the boundary of an isolating interval. In general, Definition \ref{def:weakRootIsolation} does not permit the endpoints to be roots, except for two special cases. 
	\begin{itemize}
		\item \emph{Singleton isolating intervals:} a singleton isolating interval $I_j = \{a_j\} = \{b_j\}$ contains only a root, and must be expanded into an open isolating interval without absorbing a second root. If $b_{j-1}<a_j$, then the new left endpoint of $I_j$ can be easily chosen in the interval $(b_{j-1},a_j)$. If however $b_{j-1} = a_j$, then both $a_j$ and $b_{j-1}$ must be shifted to the left by an amount less than the distance between the root $a_j$ and the unique root in $I_{j-1}$. This is done by Lines \ref{ln:leftEndpoint0} -- \ref{ln:leftEndpoint1} of Algorithm \ref{alg:stricten}. The right endpoint $b_j$ is treated in a similar manner by Lines \ref{ln:rightEndpoint0} -- \ref{ln:rightEndpoint1}. 
		\item \emph{Roots at 0 or 1:} consider the case where 0 is both a root and the left endpoint of $I_1$. This is permitted by Definition \ref{def:weakRootIsolation}, but violates Clause \ref{cl:isoIntervalEndpointsOnBoundary}. Here $I_1$ cannot be a singleton, because otherwise $I_1 = \{0\} \nsubseteq (0,1)$. In this case, $a_1$ must be shifted to the right by an amount less than the distance between $0$ and the unique root in $I_1$. This is done by Lines \ref{ln:a1Root0} -- \ref{ln:a1Root1}. If 1 is both a root and the right endpoint of $I_J$, then $b_J$ is treated in a similar manner by Lines \ref{ln:bJRoot0} -- \ref{ln:bJRoot1}. 
	\end{itemize}
	In the discussion above, the endpoints may be shifted by any amount less than the minimum root separation of the polynomial being processed. For any $p \in \mathbb{Z}[s]$ of degree $d \geq 1$, the \textsc{MinRootSep} subroutine in Algorithm \ref{alg:stricten} returns
	$$ \varepsilon:=\frac{1}{ \left\lceil d^{\frac{3d}{2}} \right\rceil \|p\|_\infty^{d-1}} \leq d^{- \frac{d+2}{2}} \|p\|^{1-d}_2, $$
	which is strictly less than the minimum root separation of $p$ by \cite[Theorem 4.6]{mignotteMathematicsComputerAlgebra1992}. The ceiling operator $\lceil \cdot \rceil$ is invoked to ensure a rational output. 
	Observe that $\varepsilon^{-1}$ grows super-exponentially with $\deg(p)$, and the length of its binary representation is $O(d \log d + d\ell)$, where $\|p\|_\infty \leq 2^\ell$. In \textsc{PolyTrace}, $p=P$ and $\deg(P) \leq MD$, where $M$ is the number of atomic propositions. In practice, the binary representation of $\varepsilon$ becomes extremely long with even a small number of atomic propositions, severly impacting the computing time. Fortunately our knowledge of $p_1,...p_M$, each of degree bounded by $D$, can be used to our advantage.   
	\begin{remark} \label{rem:rootSep}
		Let $(a,b)$ be an isolating interval for the root $s_1$ of $p  \in \R[s]$, and let $s_2 \in \{a,b\}$ be a root of $p$ on the boundary of that interval. Suppose that $p_1p_2 \mid p$, where it is known that $p_1 \in \R[s]$ has a root in $(a,b)$, and that $p_2(s_2) = 0$. Since any root of $p_1$ is a root of $p$, it follows that $s_1$ is the unique root of $p_1$ in $(a,b)$. Clearly then $|s_1 - s_2| > \textsc{MinRootSep}(p_1p_2) $. Furthermore, there can be no other roots of $p$ between $s_1$ and $s_2$. 
	\end{remark}
	Lines \ref{ln:minRootSep0}, \ref{ln:minRootSep1}, \ref{ln:minRootSep2} and \ref{ln:minRootSep3} of Algorithm \ref{alg:stricten} exploit Remark \ref{rem:rootSep}, which allows us to compute $\varepsilon$ for a polynomial of degree $2D$, rather than $MD$. This involves choosing elements from the sets $E_j := \{ i \in \{1,...,M\} \mid p_i \text{ has roots in } I_j\}$ for $1 \leq j \leq J$, which are computed by Line \ref{ln:Ej} of Algorithm \ref{alg:splineTrace}. 
	\begin{algorithm}[H]
		\caption{Converts isolating intervals $\{I_j \subset (0,1) \mid 1 \leq j \leq J \}$ for $p = \prod_{i=1}^M p_i$ into strict isolating intervals, assuming $p_1,...,p_M  \in \mathbb{Z}[s]$.} \label{alg:stricten}
		\begin{algorithmic}[1]	
			\Function{StrictIsolatingIntervals}{$p_{1:M}, I_{1:J}, E_{1:J}$} 
			\State $b_0 := 0$
			\For{$1 \leq j \leq J$} 
			\State $a_j:= \inf I_j$, $b_j := \sup I_j$
			\EndFor
			\State $a_{J+1} := 1$
			\If{$p(a_1) = 0$ \textbf{and} $a_1 = 0$} \label{ln:a1Root0}
			\State Choose $m \in E_1$
			\State Find $1 \leq i \leq M$ such that $p_i(0) = 0$ 
			\State $a_1 \leftarrow $ \Call{MinRootSep}{$p_i \times p_m$} \label{ln:minRootSep0}
			\EndIf \label{ln:a1Root1}
			\For{$1 \leq j \leq J$}
			\If{$a_j = b_j$}
			\State Choose $m \in E_j$
			\If{$a_j = b_{j-1}$} \label{ln:leftEndpoint0} \Comment{Never entered when $j=1$.}
			\State Choose $i \in E_{j-1}$
			\State $a_j,b_{j-1} \leftarrow a_j -$\Call{MinRootSep}{$p_i \times p_m$} \label{ln:minRootSep1}
			\Else
			\State $ a_j \leftarrow \frac{a_j + b_{j-1}}{2}$
			\EndIf \label{ln:leftEndpoint1}
			\If{$b_j = a_{j+1}$} \label{ln:rightEndpoint0} \Comment{Never entered when $j=J$.}
			\State Choose $i \in E_{j+1}$ 
			\State $b_j, a_{j+1} \leftarrow b_j + $ \Call{MinRootSep}{$p_i \times p_m$} \label{ln:minRootSep2}
			\Else
			\State $ b_j \leftarrow \frac{b_j + a_{j+1}}{2}$
			\EndIf \label{ln:rightEndpoint1}
			\EndIf
			\EndFor
			\If{$p(b_J) = 0$ \textbf{and} $b_J = 1$}  \label{ln:bJRoot0}
			\State Choose $m \in E_J$
			\State Find $1 \leq i \leq M$ such that $p_i(1) = 0$ 
			\State $b_J \leftarrow  1-$\Call{MinRootSep}{$p_i \times p_m$} \label{ln:minRootSep3}
			\EndIf \label{ln:bJRoot1}
			\State \Return{$(a_{1:J}, b_{1:J})$} 
			\EndFunction
		\end{algorithmic}
		\hrulefill
		\begin{algorithmic}[1]	
			\Function{MinRootSep}{$q$} \Comment{Requires $q \in \mathbb{Z}[s]$}
			\State $d := \deg(q)$
			\State \Return $1/\left(\lceil d^{\frac{3d}{2}} \rceil \cdot \|q\|_\infty^{d-1}\right)$ \label{ln:minRootSep}
			\EndFunction
		\end{algorithmic}
		
	\end{algorithm}  
	\section{Polynomial composition complexity} \label{app:polyComp}
	If $\deg(g) \leq b$, then the general form of $g \in \R[x_1,...,x_n]$ is
	$$ g(x) =  \sum_{i_1 + \hdots + i_n \leq b} a_{i_1,...,i_n} x_1^{i_1} \hdots x_n^{i_n},$$
	subject to the obvious restriction $i_1,...,i_n \geq 0$. For any $f = (f_1,...,f_n) \in \R[s]^n$,
	\begin{equation} g \circ f = \sum_{i_1 + \hdots + i_n \leq b} a_{i_1,...,i_n} f_1^{i_1} \hdots f_n^{i_n} .  \label{eq:gof} \end{equation}
	
	Regarding arithmetic complexity, if $\deg(f_i) \leq d$ for all $i$, then $\deg(f_1^{i_1} \hdots f_n^{i_n}) \leq bd$. There are ${b + n \choose n }$ summands in \eqref{eq:gof}~\cite[Lemma 8.5]{basuAlgorithmsRealAlgebraic2006}, and forming each term $f_1^{i_1} \hdots f_n^{i_n}$ requires at most $b^2d^2$ operations~\cite[Lemma 3.15]{gerhardModularAlgorithmsSymbolic2005}.
	Thus, $\deg(g \circ f) \leq bd$, and $g \circ f$ can be computed in $O\bigg(b^2d^2{b+n \choose n }\bigg)$ arithmetic operations.
	
	We can also use the polynomial norm inequalities in \cite[Chapter 3]{gerhardModularAlgorithmsSymbolic2005} to place a bound on coefficient sizes. 
	Suppose that $\|f_j \|_\infty \leq 2^\ell$ for all $1 \leq j \leq n$. Then
	$$
	\left \| \prod_{j=1}^n  f^{i_j}_j\right \|_\infty \leq  \left \| \prod_{j=1}^n  f^{i_j}_j\right \|_1 
	\leq \prod_{j=1}^n  \| f_j \|^{i_j}_1 \leq \prod_{j=1}^n  d \| f_j \|^{i_j}_\infty \leq d^n \prod_{j=1}^n 2^{\ell i_j} = d^n 2^{\ell b}. 
	$$
	If in addition $|a_{i_j}| < 2^m$ for every coefficient of $g$ in \eqref{eq:gof}, then $\| g \circ f \|_\infty \leq d^n 2^{m\ell b}$. 
\end{appendices}
\end{document}